\documentclass[10pt]{article}
\usepackage{amsmath,amsfonts,amssymb,amsthm,bbm}
\usepackage{pdfsync}
\usepackage{graphicx}
\usepackage[latin1]{inputenc}
\usepackage{hyperref}
\usepackage[all]{xy}
\usepackage{xcolor}
\usepackage{stmaryrd}
\usepackage{rotating}
\usepackage{psfrag}
\usepackage{dsfont}
\usepackage{subcaption}

\newcommand{\bit}{\begin{itemize}}
\newcommand{\eit}{\end{itemize}}
\newcommand{\bd}{\begin{description}}
\newcommand{\ed}{\end{description}}

\newcommand{\bc}{\begin{center}}
\newcommand{\ec}{\end{center}}
\newcommand{\Ref}[1]{(\ref{#1})}

\newcommand{\C}{{\mathbb C}}

\newcommand{\R}{{\mathbb R}}
\newcommand{\Z}{{\mathbb Z}}

\newcommand{\SU}{\mathrm{SU}}

\newcommand{\SL}{\mathrm{SL}}
\newcommand{\SO}{\mathrm{SO}}

\newcommand{\su}{{\mathfrak{su}}}
\newcommand{\so}{{\mathfrak{so}}}

\newcommand{\be}{\begin{equation}}
\newcommand{\ee}{\end{equation}}
\newcommand{\bea}{\begin{eqnarray}}
\newcommand{\eea}{\end{eqnarray}}
\newcommand{\bs}{\begin{subequations}}
\newcommand{\es}{\end{subequations}}
\newcommand{\nn}{\nonumber}

\newcommand{\f}{\frac}
\newcommand{\tl}{\tilde}

\newcommand{\Id}{\mathds{1}}

\newcommand{\re}{\mathrm{Re}}

\newcommand{\ra}{\rangle}

\newcommand{\bra}[1]{\langle {#1}|}
\newcommand{\ket}[1]{|{#1}\rangle}

\renewcommand{\a}{\alpha} \renewcommand{\b}{\beta} \newcommand{\g}{\gamma}
\renewcommand{\d}{\delta}  \newcommand{\eps}{\epsilon}  \newcommand{\z}{\zeta}
 \renewcommand{\th}{\theta}      \renewcommand{\l}{\lambda}
\let\m=\mu     \newcommand{\s}{\sigma}     \let\vphi=\varphi  
\let\G=\Gamma \let\D=\Delta  \let\Th=\Theta

\newcommand{\Wthree}[6]{\left(\begin{array}{ccc} #1 & #2 & #3 \\ #4 & #5 & #6 \end{array}\right)}
\newcommand{\Wfour}[9]{\left(\begin{array}{cccc} #1 & #2 & #3 & #4 \\ #5 & #6 & #7 & #8 \end{array}\right)^{(#9)}}

\usepackage[normalem]{ulem}

\begin{document}

\title{\bf SU(2) graph invariants, Regge actions and polytopes}

\author{\Large{Pietro Don\`a$^a$, Marco Fanizza$^{a,b}$, Giorgio Sarno$^{a,c}$ and Simone Speziale$^a$}
\smallskip \\ \small{$^a$ Aix Marseille Univ., Univ. de Toulon, CNRS, CPT, UMR 7332, 13288 Marseille, France} \\
\small{$^b$Universit\'a di Pisa, Largo Bruno Pontecorvo 3, 56127, Pisa, Italy} \\ 
\small{ \emph{and} \ Scuola Normale Superiore di Pisa, Piazza dei Cavalieri 7, 56127, Pisa, Italy}
 \\
\small{$^c$Universit\`a degli Studi di Torino, via Giuria 1, 10125 Torino, Italy}
}
\date{\today}

\maketitle

\begin{abstract}
\noindent 
We revisit the the large spin asymptotics of 15j symbols in terms of cosines of the 4d Euclidean Regge action, as derived by Barrett and collaborators using a saddle point approximation. We bring it closer to the perspective of area-angle Regge calculus and twisted geometries, and compute explicitly the Hessian and phase offsets. We then extend it to more general SU(2) graph invariants associated to nj-symbols. We find that saddle points still exist for special boundary configurations, and that these have a clear geometric interpretation, but there is a novelty: Configurations with two distinct saddle points admit a conformal shape-mismatch of the geometry, and the cosine asymptotic behaviour oscillates with a generalisation of the Regge action. The allowed mismatch correspond to angle-matched twisted geometries, 3d polyhedral tessellations with adjacent faces matching areas and 2d angles, but not their diagonals. 
We study these geometries, identify the relevant subsets corresponding to 3d Regge data and 4d flat polytope data, and discuss the corresponding Regge actions emerging in the asymptotics. Finally, we also provide the first numerical confirmation of the large spin asymptotics of the 15j symbol. We show that the agreement is accurate to the per cent level already at spins of order 10, and the next-to-leading order oscillates with the same frequency and same global phase. 
\end{abstract}

\tableofcontents

\section{Introduction}
A famous result by Ponzano and Regge shows that the (homogeneous) large spin limit of the SU(2) $6j$ symbol can be approximated with the cosine of the Regge action for an Euclidean tetrahedron (see e.g. \cite{Varshalovich,Schulten:1971yv}, and \cite{Aquilanti:2010sd} for a recent review). Barrett and collaborators \cite{BarrettSU2} have obtained a 4-dimensional analogue, based on (a linear combination of) SU(2) $15j$ symbols and the notion of coherent intertwiners introduced in \cite{LS}: in this case the large spin limit is related to the cosine of the Regge action for a Euclidean 4-simplex. 
In this paper we present two independent results: the first numerical confirmation of the asymptotic formula of \cite{BarrettSU2} for the $15j$ symbol, which in particular allows us to investigate properties of the higher order corrections; an analytic asymptotic formula for graph invariants associated with larger $nj$ symbols, by which we establish a new relation between these invariants and 4d polytopes.\footnote{The $15j$ and $nj$ symbols associate an SU(2)-invariant number to a given graph. This number has nothing to do with topological properties of the graph, so the naming `graph invariants' should not be confused in that sense.} Given the technical nature of the material, we give a bird's eye view of the results in the introduction, referring the reader to the various Sections for details and explicit formulas. 

Let us begin with our motivations. 
Asymptotics of graph invariants like the $6j$ symbol have many applications, and have been studied with various methods.
Extensions of the original Ponzano-Regge result include e.g. the asymptotic analysis of $9j$ symbols \cite{Haggard:2009kv}, and of $6j$ symbols for non-compact groups like SU(1,1) \cite{Davids:1998bp} and quantum groups like $\SU_q(2)$ \cite{taylor20066j}. 
More specifically, the results of \cite{BarrettSU2} were obtained following efforts in the loop quantum gravity community to find dynamical transition amplitudes in the spin foam formalism \cite{EPR,EPRL,LS2,FK}, and are closely related to similar results by the same research group for SO(4) \cite{BarrettEPRasymp} or $\SL(2, \C)$ \cite{BarrettLorAsymp} (see also \cite{Kaminski:2017eew}) invariants associated with the graph of a 4-simplex. These invariants are motivated by the EPRL spin foam model \cite{EPRL}. 
See also \cite{Haggard:2015yda} for a version of the EPRL asymptotics with the quantum group $\SL_q(2,\C)$, \cite{BCasympt1,BCasympt2,FreidelLouapre} for related asymptotic results with the previous Barrett-Crane spin foam model, and \cite{Bonzom:2011cy} for studies of inhomogeneous scalings of SU(2) $nj$ symbols.
The relevance of these results for quantum gravity is to provide a semiclassical bridge between spin foams and general relativity, albeit at the preliminary level of a single 4-simplex. Most of the results are established via an integral representation of the invariants, and a saddle point approximation in the homogeneous limit of large spins (or irreducible labels in general).
Our main goal was to provide a numerical confirmation of this approximation, and to investigate the behaviour of quantum corrections, which would play a role in the quantum gravity interpretation of these models. We present in this paper our numerical results for the SU(2) model, and in a companion paper \cite{noiLor} those for the Lorentzian EPRL model. 
The relevance of our results for spin foam models of quantum gravity will be discussed below in Section~\ref{SecQG}. For the rest of this Introduction, we focus on the results themselves, and their interest from an intrinsic algebraic and geometrical point of view. After all, SU(2) graph invariants have a wide range of applications, and thus potentially also our results on their large spin asymptotics.

At the root of Ponzano and Regge's result, there is the natural relation of the $6j$ symbol to an Euclidean tetrahedron: the $6j$ is an SU(2) graph invariant whose graph describes a tetrahedron, the spins are in 1-to-1 correspondence with the edges, and the geometry of an Euclidean tetrahedron is uniquely determined by its edge lengths. The relation of the $15j$ symbol to an Euclidean 4-simplex is less straightforward. A $15j$ symbol can indeed be associated with the graph of a 4-simplex, however its geometry is only determined by the 10 edge lengths, or equivalently the 10 triangle areas. The $15j$ symbol has thus 5 spins too many for a 1-to-1 correspondence. The key step is to take a linear combination of $15j$ symbols with the coherent intertwiners introduced in \cite{LS}. This linear combination defines an amplitude $A_v(j_{ab},\vec n_{ab})$ for ten spins and twenty unit vectors (here $a=1,..5$ labels the nodes, and the notation will be explained in detail below). When these vectors satisfy the closure conditions 
\be\label{Cintro}
C_a = \sum_{b\neq a} j_{ab} \vec n_{ab} = 0
\ee 
at each node, the data describe a classical geometry: a collection of five tetrahedra, whose areas are given by the spins, and whose dihedral angles and orientations are given by the unit vectors defining the normals to the faces. By construction, every two tetrahedra share a face with area given by the spin linking the nodes, but whose shape differs in general when determined from one tetrahedron or the adjacent one. The data describe thus a class of geometries more general than Regge geometries, which were studied and dubbed twisted geometries in \cite{twigeo}. Using saddle point techniques, it was shown in  \cite{BarrettSU2} that 
$ \lim_{\l\mapsto\infty}A_v(\l j_{ab},\vec n_{ab})$ has three different behaviours, depending on the data $(j_{ab},\vec n_{ab})$. For most general data, there are no saddles and the amplitude decays exponentially in $\l$. If the data satisfy \Ref{Cintro} and there exist five rotations $R_a\in\SO(3)$ such that
\be\label{Ointro}
R_a \vec n_{ab} = - R_b \vec n_{ba},
\ee
then the integrand admits a saddle point, and the large spin behaviour has a slower power-law decay $\lambda^{-6}$. The set of data is called a vector geometry \cite{BCasympt2}. 
Finally, if the data describe a Regge geometry, namely all shapes of the faces match and thus define a unique Euclidean 4-simplex, then the integrand admits two distinct saddle point. The power-law decay is still $\l^{-6}$, and the interference between the two points gives an oscillating behaviour with frequency determined by the Regge (boundary) action for the 4-simplex, which is the most important feature of the whole analysis.

We will rederive these results in details in Section 2, using a slightly different procedure than \cite{BarrettSU2}. 
Our procedure relies less on the bivector reconstruction theorem used in \cite{BarrettSU2}, and keeps the construction closer in spirit to the area-angle Regge calculus of \cite{DittrichSpeziale}. This allows us to keep certain geometric aspects of the saddle point analysis more manifest, and to highlight the explicit imposition of shape-matching conditions for configurations with two distinct saddles. 
We also push the analysis a bit further, and explicitly evaluate the Hessian at the saddle point, which was left only formally defined in \cite{BarrettSU2}. In particular, we determine its phase and found that the interference of the two saddle points gives precisely a cosine.

\begin{figure}[h!t]
    \centering

    \begin{subfigure}[b]{0.49\textwidth}
        \includegraphics[width=\textwidth]{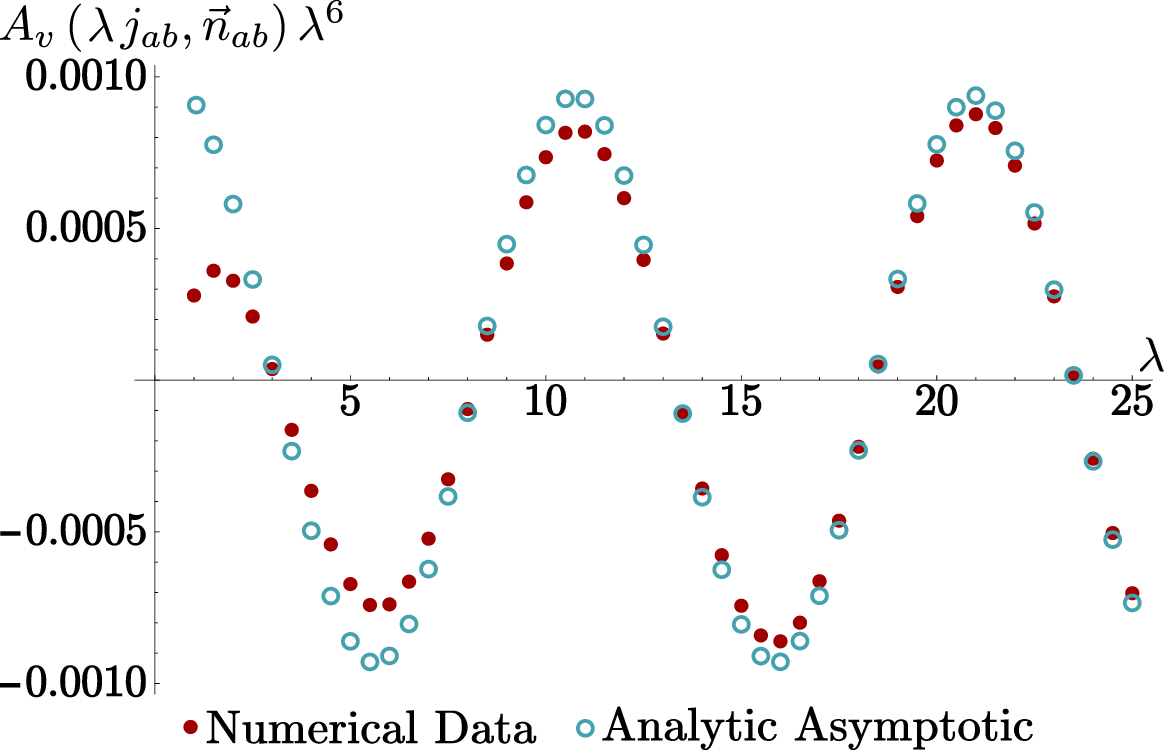}
    \end{subfigure}

\caption{\label{FigLOIntro} {\small{\emph{Numerical data versus the analytic leading order, equilateral configuration. See Section \ref{SecLO} for details.} }} }

\end{figure}

Because of the explicit dependence on SU(2) coherent states, testing numerically the asymptotics is much harder than for the $6j$ or $9j$ symbols, which possibly explains why it hasn't been done so far. The main technical step we perform is the use of recoupling theory to choose a basis of reducible $15j$ symbols, whose evaluation is much faster. Evaluating the whole amplitude is still very demanding, but becomes doable on a personal laptop at least for spins of a few tens. A detailed discussion of methods and computing times will be presented in Section~\ref{NumEv}, which contains our numerical results.  
These show excellent quantitative agreement: the asymptotic formula is remarkably accurate already at low spins, see for example  Fig.~\ref{FigLOIntro} for the equilateral configuration. An accuracy at the per cent level is reached already at spins of order 10, a situation similar to the Ponzano-Regge asymptotics of the $6j$.\footnote{The situation will be different for the Lorentzian EPRL amplitude, see \cite{noiLor}, where much higher spins will  be needed.} 

We then use our numerical data to investigate the next-to-leading order correction, and establish that it decays as one extra power of the scaling parameter, namely $\l^{-7}$, and oscillates with the same frequency as the leading order. This is to be expected on general grounds for saddle point approximations, and confirms its validity to next-to-leading order. 
This result has implications for the gravitational interpretation of these quantum corrections, which are not a modification of the action by higher order curvature invariants, but rather contribute to the measure term in the path integral. The same situation occurs in the Ponzano-Regge model \cite{Bonzom:2008xd,Kaminski:2013gaa}. The numerical results are reported in Section 3.

A question that quickly comes up when setting up numerical calculations is how to choose the orientations of the boundary data. The point is that the coherent amplitude is gauge-invariant up to a phase, due to the structure of SU(2) coherent states. Hence, it is more convenient to choose orientations to eliminate or simplify the extra phase. For Regge data, this leads us to 
consider a specific 3d object, which we refer to as \emph{twisted spike}. 
The twisted spike highlights a generic key property of the geometry at the saddle point: the 4d dihedral angles of the 4-simplex are encoded in 3d twist angles between edges of adjacent triangles. For quantum gravity models, this property is the discrete counterpart of the definition of the Ashtekar-Barbero connection in the continuum, and it shows how the Regge data satisfy the secondary simplicity constraints \cite{DittrichRyan,IoFabio,IoMiklos}.
The explicit twist angle of the twisted spike illustrates the relation between descriptions of extrinsic curvature in the covariant and canonical frameworks.
We complete our analysis by providing an explicit parametrisation of vector geometries as a subclass of twisted geometries, discussing their complete shape mismatch and description in terms of shape variables. These results are contained in Section 4, and are also relevant to the numerical analysis of the Lorentzian EPRL model, to appear in forthcoming paper \cite{noiLor}.

The asymptotic behaviour of the $15j$ symbol shows a beautiful interplay between SU(2) semiclassics and the Euclidean geometry of a 4-simplex, and it is spontaneous to ask whether there exist a similar relation between more general graph invariants and 4d geometric objects. Natural candidates are  flat convex 4d polytopes and SU(2) invariants on graphs dual to their boundary. 
The definition of the coherent amplitude $A_v(j_{ab}, \vec n_{ab})$ extends naturally to arbitrary graph invariants, where it can be expanded in a basis of $nj$ symbols, and so does the twisted geometry interpretation of generic boundary data. 
Our procedure to study the saddle point approximation extends straightforwardly, and we find the same structure: for non-degenerate configurations, generic data have no saddle points, vector geometries have one saddle point, and a special subset of them have two distinct saddle points. The special subset puts restrictions on the combinatorial structure of the graph, satisfied by graphs dual to the boundary of a 4d polytope. It spans however more general geometries than 3d polyhedral Regge geometries (which are uniquely characterised by the edge lengths): it corresponds to twisted geometries with only partial shape matching:  areas and  2d dihedral angles (namely the internal angles to the faces) match, and thus the valence of the faces; but not the diagonals, nor the edge lengths. The residual mismatch corresponds to area-preserving conformal transformations of the polygonal faces, which change their diagonals. For this reason, we can refer to such angle-matched twisted geometries also as a class of conformal twisted geometries. Our results confirm and generalise the findings of \cite{BahrSteinhaus15}, where it was shown that for a hypercube graph and regular parallelepiped data, geometries with conformal mismatches at the faces still admitted two distinct saddle points.\footnote{The result of \cite{BahrSteinhaus15} was derived for the Euclidean EPRL model, whereas we consider here SU(2) BF theory. But the differences are irrelevant for this aspect of the saddle point analysis.} 

For the angle-matched twisted geometries with two saddle points, we find that the asymptotics oscillate with frequency given by a generalised Regge action in the form
\be\label{Sintro}
S_{\G}[j,\vphi] = \sum_{ab} j_{ab} \th_{ab}(\varphi),
\ee
where $\vphi$ are the angles between the normals $\vec n_{ab}$, plus constraints imposing closure and 2d angle matching.  This boundary action has some interesting properties common to the Regge action, for instance it provides a correct discretisation of the extrinsic curvature of the 3d geometry, and it admits an extension to a 4d tessellation with proper gluing and a well-defined notion of Regge curvature. On the other hand, because of the partial shape mismatch and associate metric discontinuity, the space of solutions is larger than that of Regge calculus, and its geometric meaning and possible continuous limit are unknown to us. 

The angle-matched twisted geometries have two interesting subsets of data: 3d Regge data, and flat polytope data. 
The first can be easily characterised imposing the remaining shape-matching conditions, and corresponds to 
polyhedral Regge configurations, namely piecewise-flat geometries described by the edge lengths. In the case of the 4-simplex, the 3d Regge data identify a unique flat 4-simplex. This is not true in general, since a 3d Regge geometry on a tessellation of the 3-sphere cannot always be flatly embedded. Generic 3d Regge data correspond to a curved 4d tessellation, and do not identify a flat polytope. Notice that the asymptotic action \Ref{Sintro} cannot capture the bulk curvature, simply because there are no internal faces to define the deficit angles. Hence, 
even if the asymptotic action is now a function of edge lengths, it still fails to be a 4d Regge action in the sense that it does not describe the extrinsic curvature of a flat polytope. To achieve that, one has to further restrict the data to describe a 3d Regge tessellation that can be flatly embedded. To solve this problem, we provide an explicit criterion for flat-embedding based on an application of 4d Minkowski theorem, and the fact that the angle-matched twisted geometries can be used to define an auxiliary flat convex polytope: this identification is infinite-to-one, and the polytope is auxiliary in the sense that its areas do not coincide with the values of the spins. Requiring the areas to match the spins and imposing 4d closure select data that can be flatly embedded. 

These results, concluded with a discussion of the 4d polytope Regge action, are reported in Section 5.
They are relevant also for spin foam models, where more general graph amplitudes based on higher-valence vertices are necessary if one wants to consider arbitrary spin network states  \cite{KKL,CarloGenSF}. 
For reference's ease, we summarise in the following table the different geometric structures relevant to the saddle point analysis of SU(2) graph invariants, comparing the case of the 4-simplex and a graph dual to the boundary of a more general polytope. Here dim. is the dimensionality of the space of data, $N$ and $L$ the number of nodes and links of the general polytope graph, and $E$ is the number of edges of the boundary tessellation, which for a dominant-class polytope (in the classification of \cite{IoPoly}, namely all boundary vertices 4-valent) coincides with $2(L-N)$. 

\begin{table}[!h]
\noindent
\begin{center}
\begin{minipage}[t]{0.4\textwidth}\centering%
\vspace{-3.9cm}
\begin{tabular}{|c|c|c|}
\multicolumn{3}{c}{\bf 4-simplex graph} \\ \multicolumn{3}{c}{} \\
\hline {\bf dim.} & {\bf geometry type} & {\bf saddles} \\ && 
\\
\hline
&&\\
$20$ & twisted & 0 \\ &&\\
$15$ & vector & 1 \\ &{\small\emph{(anti-parallel)}} & \\ &&\\
10 & Regge & 2 \\ &{\small\emph{(angle-matching)}} & \\
\hline
\end{tabular}
\end{minipage}
\begin{minipage}[t]{0.4\textwidth}\centering%
\begin{tabular}{|c|c|c|}
\multicolumn{3}{c}{\bf polytope graph} \\ \multicolumn{3}{c}{} \\ \hline 
{\bf dim.} & {\bf geometry type} & {\bf saddles} \\ &&  
\\
\hline
&&\\
$5L-6N$ & twisted & 0 \\ &&\\
$3L-3N$ & vector & 1 \\ &{\small\emph{(anti-parallel)}} & \\ &&\\
 & conformal twisted & 2 \\ &{\small\emph{(angle-matching)}} & \\ &&\\
$2L-2N$ & Regge & 2 \\ &{\small\emph{(shape-matching)}} & \\ &&\\
$4N-10$ & polytope & 2 \\ &{\small\emph{(flat embedding)}} & \\ \hline
\end{tabular}
\end{minipage}
\end{center}
\caption{Classification of geometric structures relevant to the saddle point analysis of SU(2) graph invariants.\label{TablePolytopes}}
\end{table}

In Section~\ref{SecQG} we discuss some implications and applications of our results to spin foam models of quantum gravity. The final Section~\ref{SecConcl} briefly wraps our the results and some perspectives.

Throughout the paper, we follow the conventions of \cite{Varshalovich} for the recoupling theory of SU(2), and the notation of \cite{DittrichSpeziale} for the dihedral angles, generically distinguishing 4d, 3d and 2d dihedral angles respectively as $\th$'s, $\vphi$'s and $\a$'s.
The numerical calculations were mostly performed using Wolfram's Mathematica.

\section{Semiclassics of the $15j$ symbol} 

\subsection{Coherent states and coherent intertwiners} 

The key ingredient for the large spin semiclassical behaviour is the relation between SU(2) invariants and polyhedra, based on the use of coherent states and the closure condition satisfied by the invariants, which are singlet states. To make the paper self-contained and to fix our notation and convention, we provide the required background material in this Section, which can be skipped by the reader already familiar with the results of \cite{LS,IoPoly}.

The building blocks of the construction are the coherent states for an SU(2) irreducible representation $V^{(j)}$, see e.g. \cite{Perelomov}. These are defined starting from the lowest or highest weights $\ket{j,\mp j}$ and acting with a rotation that takes the vector $\hat z=(1,0,0)$ to $\vec n:=(\sin\Theta\cos\Phi, \sin\Theta\sin\Phi,\cos\Th)$, and minimize the uncertainty in the direction. The two families of states so defined are unique up to an initial arbitrary rotation along $\hat z$, which translates into a phase freedom of the coherent states. This freedom is typically fixed requiring the coherent states to define a holomorphic representation of the group. This amounts to selecting the rotation of an angle $\Th$ in the direction of $\hat z\times \vec n$.
In the fundamental $2\times 2$ representation, the associated group element gives the Hopf section of the $SU(2)\simeq S^2\times S^1$ fibration,
\be\label{Hs}
n(\z) := \f1{\sqrt{1+|\z|^2}}\left( \begin{array}{cc} 
1 & \z \\  -\bar\z & 1
\end{array}\right), \qquad \z := -\tan\f\Th2 e^{-i\Phi}
\ee
In terms of Euler angles, $\a=\Phi, \b=\Th, \g=-\Phi$, and a generic irreducible representation $V^{(j)}$ of \Ref{Hs} is expressed by the Wigner matrix 
\be
D^{(j)}(n) = e^{-i\Phi J_z} e^{-i\Th J_y} e^{i\Phi J_z}.
\ee
Accordingly, the two families of coherent states are 
\begin{align}
& \ket{j,\vec n}:=D^{(j)}(n)\ket{j,-j}, \qquad \bra{j,m}j,\vec n\ra = D^{(j)}_{m,-j}(n)= \sqrt{\left(\begin{array}{c} 2j \\ j+m \end{array}\right)}  \f{\z^{j+m}}{(1+|\z|^2)^j}, \\
&  |j,\vec n]:=D^{(j)}(n)\ket{j,j}, \qquad \bra{j,m}j,\vec n] = D^{(j)}_{m,j}(n)= \sqrt{\left(\begin{array}{c} 2j \\ j+m \end{array}\right)}  \f{(-\z)^{j-m}}{(1+|\z|^2)^j}.
\end{align}
These are related by a parity transformation\footnote{Recall that under parity, $g, \vec n, \z, (\Th, \Phi) \mapsto g^*, -\vec n, -1/\bar\z, (\pi-\Th, \pi+\Phi)$.
The ket ${\ket{j,\vec n}}$ is a complex function of the angles, a reason for which one often prefers to denote the coherent states by the complex stereographic coordinate, as in $\ket{j,\z}$.} 
and a phase shift:
\be\label{ket]}
|j,\vec n] := D^{(j)}(n)\ket{j,j} = \eps^{(j)}\overline{\ket{j,\vec n}} = (-1)^{2j_1}e^{2ij_1\phi_1} \ket{j,-\vec n},  \qquad \eps^{(j)}_{mn} = (-1)^{j-m}\d_{m,-n}.
\ee
The parity transformation is trivially related to the standard anti-linear map on SU(2), 
\be
{\cal J}\left(\begin{array}{ccc} z^0 \\ z^1 \end{array}\right) := \left(\begin{array}{ccc} -\bar z^1 \\ \bar z^0 \end{array}\right), 
\qquad {\cal J}\ket{j,\vec n} = (-1)^{2j} |j,\vec n],
\ee
which is the notation used in \cite{BarrettSU2}.
Using both families, it is possible to write SU(2) invariants completely holomorphic in the Hopf sections, such as
\be
[j,\vec n\ket{j,\vec n'}, \qquad [j,\vec n| \vec\s \ket{j,\vec n'}. 
\ee
The following formula for the scalar product will play a role below,
\be
\bra{j,\vec n_1} j,\vec n_2\ra = \left(\f{1+\vec n_1\cdot \vec n_2}{2}\right)^{2j}.
\ee

Given these states, an $n$-valent coherent intertwiner \cite{LS} is defined by the group averaging,
\be\label{LS}
|\ket{\{j_i,\vec n_i\}} := \int dg \mathop{\otimes}_{i=1}^n g\triangleright\ket{j_i,\vec n_i} \ \in \ {\rm Inv}\otimes_{i=1}^n V^{(j_i)}.
\ee
In the large spin limit, its norm is exponentially suppressed unless the closure condition $C:=\sum_i j_i \vec n_i=0$ is satisfied, in which case it scales like $j^{-3/2}$ for non-co-planar normals and $j^{-1}$ for co-planar normals. Thanks to a theorem by Minkowski, each set of data satisfying closure describes a unique convex and bound polyhedron in $\R^3$, with $j_i$ the area and $\vec n_i$ the outgoing normal of the $i$-th face. Dividing by the action of global rotations, we can characterise the intrinsic shape of the convex polyhedron in terms of its areas and  angles $\varphi_{ij}=\arccos (\vec n_i\cdot\vec n_j)$. For a pair $ij$ of adjacent faces $\vphi_{ij}$ are the (exterior) dihedral angles, and for non-adjacent faces they give the angle between the two planes to which the faces belong. An explicit algorithm for the reconstruction of the adjacency matrix of the polyhedron, as well as its volume and edge lengths, starting from the areas and normals, is presented in \cite{IoPoly}. For the special case of a tetrahedron all faces touch one another, so the adjacency matrix is trivial.

The coherent intertwiners provide an over-complete basis for the singlet space Inv$\big[\otimes_i V^{j_i}\big]$,\footnote{As the existence of exponentially suppressed norms indicates, it is also possible to restrict the family to those configurations satisfying closure only, and still have a good over-complete basis \cite{ConradyFreidel3,EteraHoloQT}. This has the advantage of reducing the number of coherent state labels by 3, from closing vectors to shape parameters only, but for the purposes of studying the saddle point approximation of graph invariants it is easier to keep the normals as labels.
}
and can be expanded in the standard basis provided by Wigner's $3jm$ symbols via recoupling theory. For $n=4$, we define the generalised Wigner $4jm$ symbol as
\begin{equation}\label{4jm}
\Wfour{j_1}{j_2}{j_3}{j_4}{m_1}{m_2}{m_3}{m_4}{j_{12}}  
= \sum_{m_{i}} (-1)^{j_{12} - m_{j_{12}}}  \Wthree{j_1}{j_2}{j_{12}}{m_1}{m_2}{m_{j_{12}}} \Wthree{j_{12}}{j_3}{j_4}{-m_{j_{12}}}{m_3}{m_4},
\end{equation}
which satisfies
\begin{align}
& \sum_{m_i} \Wfour{j_1}{j_2}{j_3}{j_4}{m_1}{m_2}{m_3}{m_4}{j_{12}} \Wfour{j_1}{j_2}{j_3}{j_4}{m_1}{m_2}{m_3}{m_4}{l_{12}} = \f{\d_{j_{12}l_{12}}}{d_{j_{12}}}, \\
& \int dg \bigotimes_{i=1}^4 D^{(j_i)}_{m_in_i}(g) =  
\sum_{j_{12}} d_{j_{12}} \Wfour{j_1}{j_2}{j_3}{j_4}{m_1}{m_2}{m_3}{m_4}{j_{12}} \Wfour{j_1}{j_2}{j_3}{j_4}{n_1}{n_2}{n_3}{n_4}{j_{12}}.
\end{align}
Then, the decomposition of a coherent intertwiner say in the recoupling channel $j_{12}$ with all links outgoing, is
\be
\label{coeffCS}
c_{j_{12}}(\vec n_i) := \bra{j_i,j_{12}}j_i, \vec n_i\ra = \sum_{m_i} \Wfour{j_1}{j_2}{j_3}{j_4}{m_1}{m_2}{m_3}{m_4}{j_{12}} 
\bra{j_i, m_i}j_i, \vec n_i \ra
=  \raisebox{-5mm}{\includegraphics[width=2.8cm]{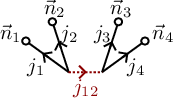}},
\ee
where in the last equality we have introduced a graphical notation that will be useful in the following.
Using the symmetries of the $4jm$ symbols, this equality is unchanged if all links are incoming. If a single link orientation is inverted, this introduces an $\eps$ tensor in the intertwiner, that can be reabsorbed in the coherent state using  \Ref{ket]}:
\begin{align*}
 \raisebox{-5mm}{\includegraphics[width=2.8cm]{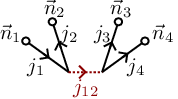}} &= \sum_{m_i} \Wfour{j_1}{j_2}{j_3}{j_4}{-m_1}{m_2}{m_3}{m_4}{j_{12}} (-1)^{j_1-m_1} \bra{j_1, \vec n_1}j_1,m_1 \ra \prod_{i\neq 1} \bra{j_i, m_i}j_i, \vec n_i \ra \\ &= \sum_{m_i} \Wfour{j_1}{j_2}{j_3}{j_4}{m_1}{m_2}{m_3}{m_4}{j_{12}} \bra{j_i,m_i} j_i,\vec n_1 ] \prod_{i\neq 1} \bra{j_i, m_i}j_i, \vec n_i \ra.
\end{align*}

These formulas can be immediately extended to arbitrary $n$. 

\subsection{Definition of the amplitude}

We consider the following $15j$ symbol,
\begin{align}\label{15j}
\{15j\} &:= \sum_{m_{ab}} (-1)^{j_{ab}-m_{ab}} \Wfour{j_{12}}{j_{13}}{j_{14}}{j_{15}}{m_{12}}{m_{13}}{m_{14}}{m_{15}}{i_1} \Wfour{j_{23}}{j_{24}}{j_{25}}{j_{12}}{m_{23}}{m_{24}}{m_{25}}{-m_{12}}{i_2}  
\\ \nn & \Wfour{j_{34}}{j_{35}}{j_{13}}{j_{23}}{m_{34}}{m_{35}}{-m_{13}}{-m_{23}}{i_3}
\Wfour{j_{45}}{j_{14}}{j_{24}}{j_{34}}{m_{45}}{-m_{14}}{-m_{24}}{-m_{34}}{i_4}
\Wfour{j_{15}}{j_{25}}{j_{35}}{j_{45}}{-m_{15}}{-m_{25}}{-m_{35}}{-m_{45}}{i_5}
\\ \nn & = \raisebox{-15mm}{\includegraphics[width=3.5cm]{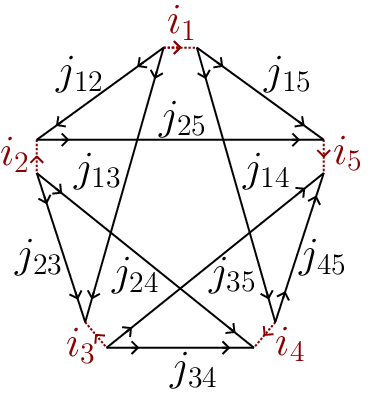}},
\end{align}
which arises as the vertex amplitude of 4d simplicial quantum SU(2) BF theory \cite{Ooguri:1992eb}. The generalised Wigner's $4jm$ symbols used above were defined in \Ref{4jm}.
Like all recoupling symbols it depends on the orientation of the links. Here we chose the one depicted in the graphical representation above. With this choice of orientation, we can number the nodes so that the direction of a link $ab$ is always from $a<b$ to $b$. Reversing a line carrying spin $j$ results in an overall phase $(-1)^{2j}$.

The interest in this SU(2) invariant is that its graph is in correspondence with (the dual to) the boundary of a 4-simplex, where each 4-valent node, dual to a tetrahedron, has been split into two 3-valent ones.
Thus, the recoupling scheme distinguishes a set of 5 spins $i_a$, $a=1,\ldots 5$, that we will refer to as intertwiner spins. 

The actual asymptotic we are interested in involves not a single $15j$, but a special linear superposition, which allows us to endow this fundamentally 3d Euclidean object with a 4d geometric interpretation. To that end, we project \Ref{15j} on coherent intertwiners \cite{LS}, built from group averaging SU(2) coherent states $\ket{j,\vec n}$ as described in the previous Section. 
To define the coherent intertwiners, we need 4 unit vectors per node. This results in the assignment of two vectors per link, which we denote $\vec n_{ab}$ and $\vec n_{ba}$. It is also convenient to use the graph orientation to keep track of outgoing and incoming links at each node, and include a parity transformation when assigning vectors to say the incoming links. Accordingly, we associate a ket $\ket{j_{ab},\vec n_{ab}}$ to each outgoing link, and a parity-reversed bra $\bra{j_{ab},-\vec n_{ba}}$ to each incoming one. This rule leads to a simplification of the saddle point analysis, as we will explain below: it makes the parallel transport of two opposite normals the identity, instead of the parity transformation.

The resulting coherent vertex amplitude can be written as
\begin{equation}\label{Av}
\begin{split}
A_v(j_{ab},\vec n_{ab})= 
\int \prod_a dg_{a} \prod_{(ab)}\bra{-\vec{n}_{ab}}g_{a}^{-1}g_{b} \ket{\vec {n}_{ba}}^{2j_{ab}} 
= \sum_{\{i_a\}} \prod_{a} d_{i_a}
\raisebox{-20mm}{\includegraphics[width=4.5cm]{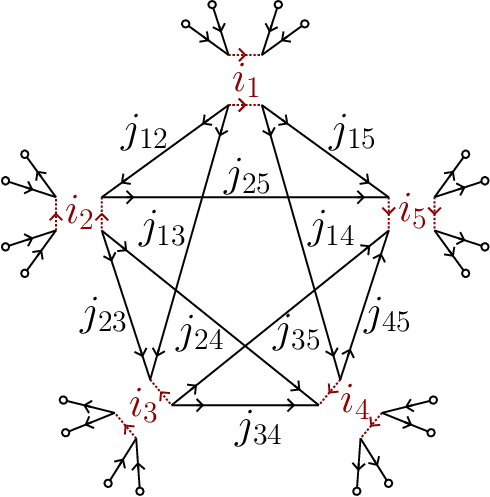}},
\end{split}
\end{equation}
where the product over $(ab)$ means over the oriented links, 
and $\ket{\vec n_{ab}}:=\ket{\tfrac12, \vec n_{ab}}$ is the coherent state in the fundamental representation. 
This is the quantity whose large spin behavior we are interested in. 
The second equality above provides its expression as linear combinations of the $15j$ symbols weighted by coherent intertwiners and dimensional factors. Notice that using recoupling theory, we could have equally decomposed \Ref{Av} on a basis of $15j$ symbols of any other kind, not just \Ref{15j}, a freedom that will play an important role in our numerical investigations. 

An alternative definition, used in \cite{BarrettSU2}, is to put the parity-inversion at the level of the ket instead of the classical label, and assign the dual bras $[j_{ab},\vec n_{ab}|$ to the target nodes. We see from \Ref{ket]}, that these two definitions are related by a global phase:
\begin{align}\label{Aholo}
A_{\rm holo}(j_{ab},\vec n_{ab}) &= \int \prod_a dg_{a} \prod_{(ab)}[\vec{n}_{ab}| g_{a}^{-1}g_{b} \ket{\vec {n}_{ba}}^{2j_{ab}} 
= \sum_{\{i_a\}} \{15j\} \prod_a  d_{i_a} c_{i_a}(n_{ab})  \\\nonumber&
= e^{-i\sum_{ab} j_{ab}\Phi_{ab}} A_v(j_{ab},\vec n_{ab}). 
\end{align}
The graphical notation is the same as above with now all coherent intertwiners outgoing as in \Ref{coeffCS}.
This alternative has the advantage that the amplitude is holomorphic in all the Hopf sections \Ref{Hs} of the coherent states. Furthermore, only the intertwiner with all outgoing links \Ref{coeffCS} is required in the expansion in terms of $15j$ symbols.
On the other hand, the saddle point analysis is clearer in the first version \Ref{Av} of the amplitude: the extra global phase of \Ref{Aholo} adds nothing to the Regge-like large spin limit. In fact, it should be clear from the discussion above that the global phase of the coherent vertex amplitude does not capture any property of the $15j$ symbol, since it can be changed arbitrarily changing the phase definition of the SU(2) coherent states.

A further remark concerning phases.
We follow the conventions of \cite{Varshalovich}\footnote{These are mostly the same as those of Wolfram's Mathematica, one notable exception being the Wigner matrices, one the inverse of the other.}, in particular all Clebsch-Gordan coefficients are real and thus also the recoupling symbols. On the other hand, the coherent states are complex, and so are the amplitudes \Ref{Av} and \Ref{Aholo} as well as their large spin asymptotics. 
However, only the absolute value of \Ref{Av} is invariant under SU(2) transformations at the nodes, the phase is not. The whole coherent amplitude is thus only SU(2) covariant, and one can use the freedom of changing the phase of the coherent states to make \Ref{Av} real, as argued in \cite{BarrettSU2} and reviewed below.

\subsection{The boundary data: twisted, vector and Regge geometries}\label{sectwi}

Before reviewing the saddle point approximation of \Ref{Av}, let us recall the geometric interpretation of the boundary data. 
This is given by a set of ten spins $j_{ab}=j_{ba}$, one per link, and twenty unit vectors $\vec n_{ab}\neq \vec n_{ba}$, two per link, for a total of $50$ variables. 
The coherent vertex amplitude is  invariant under SU(2) transformations at the nodes: information on the local orientation of the normals is lost, and the amplitude depends only on the angles among them. These are five linearly independent ones per node, $\vec n_{ab}\cdot \vec n_{ac}=\cos\vphi^{(a)}_{bc}$, hence the amplitude depends only on $10+5\times 5=35$ variables. Among those, configurations not satisfying the closure conditions are expected to be exponentially suppressed by the result of \cite{LS}, and indeed this was confirmed in \cite{BarrettSU2}. Consider then only the closed boundary data. We restrict attention to non-coplanar normals at each node. These identify tetrahedra, and $\vphi$ are their dihedral angles
The gauge-invariant set has $35-3\times 5=20$ variables, one area per link and two shape variables per tetrahedron. It thus describes a collection of 5 tetrahedra attached by the triangles, where each triangle has a unique area, but two different shapes determined by the two tetrahedra: a twisted geometry \cite{twigeo}.\footnote{\label{footxi} The most general twisted geometry, defined by a symplectomorphism to gauge-invariant SU(2) holonomies and fluxes \cite{twigeo} (see also \cite{twigeo2,IoPoly,EteraTamboSpinor,IoFabio,IoMiklos}), carries also an additional angle per link. This angle, denoted $\bar\xi_{ab}$ in the gauge-invariant version discussed in \cite{IoMiklos}, describes the twist between edges, and can be used to encode the classical extrinsic curvature between two polyhedra. This additional label does not play any role here, because we are interested in asymptotics at fixed spins, namely with sharp areas that we identify as the (discrete) classical areas. The extrinsic curvature is reconstructed in the amplitude from the local flat embeddings guaranteed by the $g_a^{-1}g_b$ structure in its definition.} If we require all the shapes to match, this introduces 10 independent conditions, thus $20-10=10$ independent variables, that can be identified with the areas. 
The resulting subspace is in 1-to-1 correspondence with a flat 4-simplex. Depending on the values of areas and normals, the 4-simplex can either be Euclidean, or Lorentzian with all boundary tetrahedra space-like, see below Section \ref{SecCritical} for more details. 

The asymptotic behavior of \Ref{Av} distinguishes two further subclasses of data: vector geometries and Regge geometries.
The results of \cite{BarrettSU2} show that, for general boundary data, \Ref{Av} decreases exponentially as the ten spins are homogeneously rescaled. An enhancement occurs only for measure-zero special set of data, for which the integrand has a saddle point and the amplitude decreases only as a power-law. This special set of data is called a vector geometry and corresponds to a 15-dimensional subset of the twisted geometries, for which closure holds and furthermore the normals can be made pairwise anti-parallel with rotations of each tetrahedron.
For the interested reader, we will provide below in Section \ref{SecVecGeo} an explicit parametrization of vector geometries as a subset of twisted geometries. If we further restrict to the 10-dimensional subset of Regge geometries, there exist a second saddle point, and the relative phase between the two saddle points is proportional to the Regge action. Since the global phase of the amplitude can be arbitrarily changed by re-phasing the coherent states, it is only for configurations with two saddles that one can really claim an interesting semiclassical limit of the $15j$ symbol.

\subsection{Critical points}\label{SecCritical}
We want to estimate the integral \Ref{Av}  in the asymptotic limit with all spins uniformly large. To that end, we redefine the spins as $j_{ab} \rightarrow \lambda j_{ab}$, and rewrite the integrands in exponential form, 
\be
\label{15SU2action}
A_v(j_{ab},\vec n_{ab})=\int \prod_{a } dg_{a} \, e^{\l S(g_a;j_{ab},\vec n_{ab})}, 
\qquad S(g_a;j_{ab},\vec n_{ab})=\sum_{1 \le a < b \le 5}2j_{ab}\log\bra{-\vec{n}_{ab}}g_{a}^{\dagger}g_{b}\ket{\vec{n}_{ba}}. 
\ee
Being the action complex, a saddle point approximation for $\lambda \rightarrow \infty$ is usually complicated. A general strategy is to look for critical points at which the real part of the gradient of the action vanishes, then deform the integration contour to the complex plane to make the phase stationary. And if more than one such critical points exist, one selects the one with the largest real part of the action. On the other hand, the case at hand is substantially easier: first, since
)
\be
|\bra{j_{ab},-\vec{n}_{ab}} g_{a}^{\dagger}g_{b} \ket{j_{ab}, \vec{n}_{ba}}|^2 = \left(\frac{1-\vec{n}_{ab}\cdot R_{a}^{-1}R_{b}\vec{n}_{ba}}{2}\right)^{2j_{ab}},
\ee
where $R:=D^{(1)}(g)$ are SU(2) matrices in the adjoint representations,
the real part of the action is always negative, and its absolute maxima $\re S=0$ easily identified at values of the rotations making the vectors anti-parallel:
\begin{subequations}\label{CP}\be\label{CP1}
R_{b}\vec{n}_{ba}=-R_{a}\vec{n}_{ab};
\ee
second, as we will review below, the imaginary part of the gradient can be simultaneously made to vanish, if the boundary data satisfying the closure conditions
\be
\sum_{b\neq a}j_{ab}\vec{n}_{ab} = 0 \quad \forall a.
\ee
\end{subequations}
Hence, critical points exist only for boundary data satisfying closure and whose normals can be made pairwise anti-parallel with local rotations at the nodes. Following Barrett \cite{BCasympt2} we will refer to such data as vector geometries, and to \Ref{CP1} as orientation equations. The critical points themselves are the values of the rotations realising \Ref{CP1}, and depend on the configuration of normals.
To find them, let us first consider the special set of boundary data with all normals pairwise opposite,
\be
\label{vecgeo}
\vec n_{ab} = -\vec n_{ba}.
\ee 
For such boundary data, $R_a=\Id$ is of course a solution of \Ref{CP1}. The question is whether this solution is unique or not.
To address this question, let us use the global SU(2) symmetry of \Ref{15SU2action} to eliminate one of the integration variables and fix say $g_1=\mathds{1}$. We then split the ten vectorial equations \Ref{CP1} into the four involving  $R_1$ and the remaining six, 
\begin{subequations}\label{rotCritPoint}\begin{align}
& R_{b}\vec{n}_{b1}=-\vec{n}_{1b} = \vec{n}_{b1},  \label{rotCritPoint1} \\
& R_{a}^{-1} R_{b}\vec{n}_{ba} =-\vec n_{ab} = \vec{n}_{ba}, \label{rotCritPoint2}
\end{align}\end{subequations}
where in the second equalities we used our assumption \Ref{vecgeo} on the boundary data.
Let us look first at \Ref{rotCritPoint1}: these equations are solved by a rotation around $\vec{n}_{b1}$ of an arbitrary angle, which for later convenience we denote $2\th_{1b}$:
\be\label{Rpar}
R_b := e^{i2\th_{1b} \vec n_{b1}\cdot \vec J}.
\ee
To determine the angle, we look at \Ref{rotCritPoint2}, with $a,b\neq 1$.
Using \Ref{Rpar} and the familiar composition of rotations, we get $R_{a}^{-1} R_{b}=e^{2i\th_{ab} \vec u_{ba}\cdot \vec J}$, a rotation 
of an angle $2\theta_{ab}$ such that
\be
\label{sphericalcosinelaw}
\cos\theta_{ab}	=\cos\theta_{1b}\cos\theta_{1a}+\sin\theta_{1b}\sin\theta_{1a} \, \vec{n}_{1a}\cdot\vec{n}_{1b},
\ee 
around the axis
\be\label{axis}
\sin\th_{ab}\, \vec u_{ba} = -\sin\theta_{1a}\cos\theta_{1b} \, \vec{n}_{a1}+\cos\theta_{1a}\sin\theta_{1b} \, \vec{n}_{b1}-\sin\theta_{1a}\sin\theta_{1b} \,\vec{n}_{a1}\times\vec{n}_{b1}.
\ee
Then, the critical point equations \Ref{rotCritPoint2} require $\vec u_{ba}=\vec{n}_{ba}$. This gives us two equations which fix $\th_{1a}$ and $\th_{1b}$: the scalar product of \Ref{axis} with $\vec{n}_{a1}\times\vec{n}_{ba}$ must vanish, and assuming $\th_{1a}\neq 0$, we obtain
\be
\label{angleSolution}
\tan\theta_{1a}= 
\frac{\vec{n}_{1b}\cdot\vec{n}_{1a}\times\vec{n}_{ba}}{\vec{n}_{b1}\cdot\vec{n}_{ba}+\vec{n}_{a1}\cdot\vec{n}_{ab} \, \vec{n}_{1b}\cdot\vec{n}_{1a}};
\ee 
and similarly for $\th_{1b}$.
If on the other hand $\th_{1a}=0$, then it must be so for all $a$ as
a direct consequence of \Ref{rotCritPoint2},
and from \Ref{sphericalcosinelaw} we see that also $\th_{ab}=0$: we recover in this way only the trivial solution with all rotations being the identity. 

Now let us scrutinise the non-trivial solution more carefully: 
a little trigonometry shows that the equation above is equivalent to
\be\label{thtophi}
\cos\th_{1a} = \cos\th_{1a}^{(b)}(\vphi) := \f{\cos\vphi^{(b)}_{1a} + \cos\vphi^{(a)}_{1b} \cos\vphi^{(1)}_{ab}}{\sin\vphi^{(a)}_{1b} \sin\vphi^{(1)}_{ab}},
\ee
which is the spherical cosine law between 4d and 3d dihedral angles of a flat 4-simplex, respectively $\th_{1a}$ and $\vphi^{(1)}_{ab} = \arccos (\vec n_{1a}\cdot\vec n_{1b})$, etc. 
On the other hand,  \Ref{angleSolution} and thus \Ref{thtophi} must hold for arbitrary $a$ and $b$. Since the left-hand side does not depend on the choice of $b$, non-trivial solutions exist only for those configurations for which the right-hand side also does not depend on the choice of $b$. This gives us the following edge-independence conditions, 
\be\label{edge-ind}
{\cal C}_{1a,bc} = \f{\cos\vphi^{(b)}_{1a} + \cos\vphi^{(a)}_{1b} \cos\vphi^{(1)}_{ab}}{\sin\vphi^{(a)}_{1b} \sin\vphi^{(1)}_{ab}}
-\f{\cos\vphi^{(c)}_{1a} + \cos\vphi^{(a)}_{1c} \cos\vphi^{(1)}_{ac}}{\sin\vphi^{(a)}_{1c} \sin\vphi^{(1)}_{ac}}=0,
\ee
which assure that the value of the 4d dihedral angle $\th_{1a}$ does not depend on the edge chosen to compute it with the spherical cosine law. 
Notice also that substituting the solutions \Ref{thtophi} in \Ref{sphericalcosinelaw}, we derive a spherical cosine law for the angles $\th_{ab}$ of the rotations between the node $a$ and $b$, hence $\th_{ab}$ are also 4d dihedral angles. The substitution can also be used to show that the edge-independence conditions \Ref{edge-ind} hold with an arbitrary node $d$ replacing $1$.\footnote{Explicitly, one can first write the spherical cosine law for $\th_{1a}$ with $b$ as third node, and for $\th_{1b}$ with $a$ as third node. Plugging these in \Ref{sphericalcosinelaw} gives a spherical cosine law for $\th_{ab}$ with 1 as the third node. Choosing different third nodes for $\th_{1a}$ and $\th_{1b}$ will give more complicated expressions for $\th_{ab}$, and their equality implied by \Ref{edge-ind} implies also edge-independence conditions with arbitrary $d$ replacing 1.}
We have excluded from the analysis the case in which the normals at one node are coplanar, for which the 3d volume and one or more of the $\vphi$ angles are zero, making the spherical cosine laws and 3d geometry ill-defined. 

At this point we have determined all the group elements at the second saddle point, and established that its existence requires data satisfying the edge-independence conditions. It is easy to show (see for instance \cite{DittrichRyan} and below in Section~\ref{SecCTG}) that these edge-independence conditions imply 
the shape-matching conditions of \cite{DittrichSpeziale}, namely adjacent tetrahedra define shared triangles of the same shape. Hence, areas and 3d angles admitting a second saddle point determine a unique Euclidean 4-simplex.\footnote{Using shape-matched areas and 3d angles also avoids the issue of special regular configurations for which the Jacobian between edge lengths and areas vanishes, notably the case with all equal areas, to which both an equilateral and a Tuckey configuration with 7 equilateral and 3 isosceles triangles exist \cite{BCasympt2}.
Notice that shape-matched areas and 3d angles can also determine a unique Lorentzian 4-simplex, with all tetrahedra space-like. For those configurations however, the spherical cosine laws \Ref{thtophi} determine hyperbolic cosines instead of cosines, and \Ref{CP1} is not satisfied. Hence the SU(2) coherent amplitude \Ref{Av} has no critical points and its asymptotic behaviour exponentially suppressed. Lorentzian configurations are on the other hand relevant for the Lorentzian EPRL model, see \cite{BarrettLorAsymp} and our companion paper \cite{noiLor}.}

Summarising, pairwise-antiparallel normals satisfying closure admit a critical point $R_a^{(0)}$ at the identity; if they additionally satisfy the shape-matching conditions, a second critical point $R_a^{(\th)}$ exists, for which the rotation at a chosen node is still the identity, and all others are a rotation by (twice) the 4d dihedral angle in the direction of the normal outgoing to the chosen node. Notice that to derive this result we used the vectorial representation of SU(2). When going back to the fundamental representation which defines our integral, both candidate saddle points have a two-fold degeneracy due to the double covering of SU(2) on the group of rotations:
\be\label{double}
R_a^{(0)}\equiv \Id_3 \mapsto g_a^{(0,\pm)}\equiv \pm\Id_2, \qquad R_a^{(\th) } \mapsto g_a^{(\th,\pm)}= \pm g_a^{(\th)}.
\ee
This sign degeneracy is nothing but the usual $\Z_2$ symmetry in the map between SU(2) spinors and vectors, and plays no role in the evaluation of the integral: changing any one of the signs at the saddle point we get a factor
$(-1)^{2\sum_b j_{ab}}$, which is always unit because of the Clebsch-Gordan conditions at the node. 
For this reason, we will still refer in the following to configurations admitting a single or two critical points, instead of 2 and 4. The reader should keep in mind the vectorial representation when we say so.

\subsection{Expansion around the critical points}
In the previous subsection we found two saddle points, $g_a^{(0)}$ and $g_a^{(\th)}$, at which the real part of the action has an absolute maximum, and thus the real part of the gradient vanishes, since the action is a periodic function. On the other hand, we have not said anything yet about the imaginary part of the gradient, so we do not know whether these points can be used for a good approximation of the integral. To answer this question, let us Taylor expand the action around $g_a^{(\th)}$; the expansion around the trivial point $g_a^{(0)}$ can be straightforwardly obtained for $\th_{ab}=0$. 
To do so we parametrise   
\be
g_{a}=g_{a}^{(\th)}\exp\left(\f i2 \vec{m}_{a}\cdot \vec{\sigma}\right) = g_{a}^{(\th)}\left(\mathds{1}+ \f i2 \vec{m}_{a}\cdot \vec{\sigma}-\frac{1}{8}\vec{m}_{a}\cdot\vec{m}_{a}\mathds{1}+o(|\vec m_a|^3)\right)
\ee
for $a=2,\ldots 5$, where we assumed  $\left|\vec{m}_{a}\right|\ll1$.
Inserting these expansions in the action \Ref{15SU2action} we obtain
\begin{align}\nn
S &=\sum_{(ab)}2j_{ab}\log\bra{-\vec{n}_{ab}}\left(\mathds{1}-\f i2 \vec{m}_{a}\cdot \vec{\sigma} -\frac{1}{8}\vec{m}_{a}\cdot\vec{m}_{a}\mathds{1}\right)g_{a}^{(\th)}{}^{\dagger}g_{b}^{(\th)}\left(\mathds{1}+\frac i2 \vec{m}_{b}\cdot \vec{\sigma}-\frac{1}{8}\vec{m}_{b}\cdot\vec{m}_{b}\mathds{1}\right)\ket{\vec{n}_{ba}} +o(|\vec m_a|^3) \\ \nn
&=2i\sum_{(ab)}j_{ab}\theta_{ab}
+i\sum_{a=2}^{5}\vec{m}_{a} \sum_{b\neq a}j_{ab}\vec{n}_{ab}
-\frac{1}{4}\sum_{a=2}^5\sum_{b \neq a}j_{ab}\left(\vec{m}_{a}\cdot\vec{m}_{a}-\vec{m_{a}}\cdot\vec{n}_{ba} \, \vec{m_{a}}\cdot\vec{n}_{ba}\right)\\
&\hspace{1.5cm} +\frac{1}{2}\sum_{2 \le a < b \le 5}j_{ab}\exp\left(-2i\theta_{ab}\right)\left(\vec{m}_{a}\cdot\vec{m}_{b}-\vec{m}_{a}\cdot\vec{n}_{ba} \,\vec{m}_{b}\cdot\vec{n}_{ba}+i\vec{m}_{a}\times\vec{m}_{b}\cdot\vec{n}_{ba}\right) +o(|\vec m_a|^3),\label{Sexp}
\end{align}
where $\th_{ab}=\th_{ab}(\vphi)$ via \Ref{thtophi}.

The gradient is purely imaginary, as anticipated at the beginning of this Section, and furthermore can be made to vanish if the closure conditions \Ref{CP1} are satisfied for every tetrahedron.\footnote{What \Ref{Sexp} actually imposes is the closure at the four tetrahedra $a=2\ldots 5$, but the closure for tetrahedron 1 then follows from the vector geometry conditions \Ref{vecgeo}.} The zeroth-order action is also purely imaginary, with 
\be\label{imS}
-\f i2 S(g_{a}^{(\th)};j_a,\vec n_{ab}) = S_{\rm R}:=\sum_{(ab)} j_{ab}\theta_{ab}(\vphi).
\ee
It can be identified with the Regge boundary action for a 4-simplex because, although a priori the areas $j_{ab}$ and dihedral angles $\vphi^{(a)}_{bc}$ are independent variables, the non-trivial saddle point $g_{a}^{(\th)}$ only exists provided that the closure and shape-matching conditions are satisfied.\footnote{We stress that when we talk about Regge action we are making a statement about the independent variables to be varied, and not merely about the apparent form of the action. If it weren't for closure and shape matchings being implemented, \Ref{imS} would not have much to do with the Regge action. This important point appears to be overlooked in some literature, where formal variations or resummations of the spins are done at the saddle point treating them as if the were independent -- which they are not because of the constraints. 
We will have the occasion to appreciate the difference when discussing the generalisation to polytopes below in Section \ref{SecReggePoly}. } 
It is then more appropriate in our opinion to speak of an angle-area Regge action, and think of \Ref{imS} as the following constrained action:
\be
S[j_{ab}, \vphi^{(a)}_{bc}] = \sum j_{ab} \th_{ab}(\vphi) + \sum \l_{a,b} C_{a,b}(j,\vphi) + \sum \m_{ab,cd} \, {\cal C} _{ab,cd}(\vphi),
\ee
where
\be\label{phiclosure}
C_{a,b}(j,\vphi):= j_{ab} + \sum_{c\neq a,b} j_{ac} \cos\vphi_{bc}^{(a)} = 0
\ee
is the closure constraint written in terms of angles, and the shape-matching constraints are given in \Ref{edge-ind}.
The equivalence of this action to the Regge action based on edge lengths as independent variables was proved in \cite{DittrichSpeziale}.

Finally, let us comment on the Hessian of \Ref{Sexp}. 
It is a $12\times 12$ matrix, that can be arranged in $3\times3$ blocks defined by 
\begin{align}\label{Hessian}
H_{ab}&=\frac{\delta^2 S}{\delta \vec{m}_a \delta \vec{m}_b} \\\nn& = -\frac{1}{2}\delta_{ab}\sum_{c\neq a}j_{ac}\left(\mathds{1}-\vec{n}_{ac}\otimes\vec{n}_{ac}\right) +\frac{1}{2}j_{ab}\left(1-\delta_{ab}\right)\exp\left(-2i\theta_{ab}\right)\left(\mathds{1}-\vec{n}_{ab}\otimes\vec{n}_{ab}-i\star\!\vec{n}_{ab}\right),
\end{align}
with $a,b=2\ldots 5$, consistently with the gauge-fixing chosen, and where we used a Hodge star notation for the mapping of 3d vectors to $3\times 3$ antisymmetric matrices, $(\star \vec n)^{ij}:=\eps^{ijk} n^k$.
The diagonal blocks $a=b$ given by the first term in the expression can be recognised as the Hessian arising from the coherent intertwiners \cite{LS}, and it has maximal rank for non-coplanar normals. The non-trivial 4-simplex structure is in the non-diagonal blocks mixing different tetrahedra, and which carry the dependence on $\th_{ab}$. The structure of the non-diagonal blocks is complicated enough to prevent an explicit calculation of the determinant in a compact analytic form. For our purposes, we limited ourselves to numerical studies of various configurations. We found that the determinant is non-vanishing for generic configurations,\footnote{It will vanish for degenerate configurations, e.g. the normals in a tetrahedron are all co-planar. In this case the amplitude can have a slower power law decay. These case are excluded from our analysis.} consistently with the assumption of \cite{BarrettSU2}; furthermore, the following reality property holds,
\be\label{daje}
\det - H^{(0)} = \overline{\det - H^{(\th)}}.
\ee 
To give two explicit examples: for the equilateral configuration with all spins equal 1 we have
\be
\det - H^{(0)} = \frac{1618200- i 316712 \sqrt{15}}{177147}, \qquad 
\det - H^{(\th)} = \frac{1618200+ i 316712 \sqrt{15}}{177147};
\ee
for an isosceles  configuration, with an equilateral tetrahedron with spins equal 2 and four isosceles tetrahedra of spins 2 and 1,
we get
\be\label{isosceles}
\det - H^{(0)} = \frac{158279364 - i 22307203 \sqrt{6}}{708588}, \qquad 
\det - H^{(\th)} = \frac{158279364 + i 22307203 \sqrt{6}}{708588}.
\ee
We checked that \Ref{daje} holds for more general configurations as well without symmetries, but we were not able to prove it analytically for lack of an explicit formula for the determinant. It could be proved also without computing explicitly the determinant if one could show that the Hessian matrices of \Ref{Sexp} around the two critical points are complex conjugated up to a similarity transformation. In fact, by looking directly at the action one could also prove a more general result, namely that the phase of the \emph{whole} expansion \Ref{Gaussian1} coincides with the one of the leading order, if the same similarity conjugation holds for all terms in the expansion of the action. This may be unexpected, but our numerical studies reported below suggest that it may well be the case.
To that end, we remark here that the action to all orders around the pairwise anti-parallel configuration can be written in a simple close form as
\begin{align} \label{Sexact}
S(g_{a}^{(\th)};j_a,\vec n_{ab}) &=2i\sum_{(ab)} j_{ab} \theta_{ab}
+2i\sum_{a=2}^{5} j_{ab} \log f(g_{a}^{(\th)},\vec n_{ab}), \\ \nn 
&  f(g_{a}^{(\th)},\vec n_{ab}) :=  \cos\left(\frac{|m_a|}{2}\right)\cos\left(\frac{|m_b|}{2}\right)- \frac 1 2 \vec{m}_{a}\cdot\vec{n}_{ab} \, \vec{m}_{b}\cdot\vec{n}_{ab} \nn
\\&
\qquad -e^{-i2\theta_{ab}}\left(\sin\left(\frac{|m_a|}{2}\right)\sin\left(\frac{|m_b|}{2}\right)\vec{m}_{a}\cdot\vec{m}_b+\vec{m}_{a}\cdot\vec{n}_{ab} \, \vec{m}_{b}\cdot\vec{n}_{ab}-i\vec{n}_{ab}\cdot \vec m_a \times \vec m_b\right). \nn
\end{align}
\subsection{Asymptotic formula}
We have identified proper critical points for boundary data satisfying closure and pairwise anti-parallel normals \Ref{vecgeo} and we computed the quadratic expansion of the action around them. With these results we can proceed to a Gaussian approximation of the integral. Taking into account the two-fold degeneracy of the critical points and the redundancy of one group integral, we get
\begin{align}
\label{Gaussian1}
A_v(j_{ab},\vec n_{ab})&= \nn
\sum {}_{g_a^{\rm (c)}} \ \exp\big(\l S(g_a^{\rm (c)};j_a,\vec n_{ab})\big) \int \prod_{a=1}^4 \f{d^3\vec m_a}{(4\pi)^2} \, \exp\Big(\f \l2 \sum_{a,b} \vec m_a \cdot H_{ab} \vec m_b\Big) +o(\l^{-7}) \\
&= 2^4 \sum {}_{R_a^{\rm (c)}}\ \left( \frac{2 \pi}{ \lambda} \right)^6 \frac{1}{(4 \pi)^8} \f{\exp\left(\l S(R_a^{\rm (c)};j_a,\vec n_{ab})\right) }{\sqrt{\det(-H^{\rm (c)})}}  +O(\l^{-7}).
\end{align}
For the explicit expression, we have to distinguish the two cases of shape-matched and non-shape-matched configurations.

\begin{itemize}

\item Vector geometries \Ref{vecgeo}, not satisfying the shape-matching conditions: a single critical point at the identity, and 
\begin{align}\label{LO}
A_v(j_{ab},\vec n_{ab})=\left( \frac{2 \pi}{ \lambda} \right)^6 \frac{2^4}{(4 \pi)^8} \frac{1}{\sqrt{\det - H^{(0)}}} +O(\l^{-7}). 
\end{align}

\item Regge geometries, satisfying the shape-matching conditions: two distinct saddle points, and
\begin{align}\nn
A_v(j_{ab},\vec n_{ab})&=\left( \frac{2 \pi}{ \lambda} \right)^6 \frac{2^4}{(4 \pi)^8} \left( \frac{1}{\sqrt{\det - H^{(0)}}} + \frac{e^{i 2 \lambda S_R }}{\sqrt{\det -H^{(\th)}}}\right) +O(\l^{-7})=\\\label{LO2}
&=\left( \frac{2 \pi}{ \lambda} \right)^6 \frac{2^4}{(4 \pi)^8}  \frac{e^{i \lambda S_R }}{\sqrt{\left| \det -H^{(0)}\right|}} \cos\left( \lambda S_R -\frac{1}{2}\mathrm{arg}\det-H^{(0)} \right)+O(\l^{-7}),
\end{align}
where in the second step we used \Ref{daje}.
\end{itemize}

Let us comment on the global phases of \Ref{LO} and \Ref{LO2}. With our definition \Ref{Av} of the coherent vertex amplitude, the leading order for pairwise anti-parallel boundary data \Ref{vecgeo} is automatically real, since the critical point is at the identity. For Regge boundary data, the leading order can be made real redefining the phase of the coherent states, see discussion above \Ref{Hs}. What we need in this case is 
\be
\ket{j_{ab},\vec n_{ab}} \mapsto e^{ij_{ab}\th_{ab}(\vphi)} \ket{j_{ab},\vec n_{ab}}.
\ee 
This re-phasing can harmlessly be done (although the new coherent states will not provide a holomorphic representation of the SU(2) algebra), notice however that it requires each coherent state to be redefined with a phase that depends on all the spins of the graph. 

\vspace{1em}
The result generalises easily to arbitrary critical points, i.e. configurations satisfying \Ref{CP1} but not directly \Ref{vecgeo}. 
In fact, we can use the invariance of the amplitude under SU(2) transformations at the nodes to bring us back to the case \Ref{vecgeo} studied above, and the critical points will be shifted accordingly. Denote ${R}'_{a}=\exp (i(\psi_a/2)\vec v_a\cdot \vec\s ) \in SO(3)$ the solution of \Ref{CP1}, and define $\vec{n}'_{ab}={R}'_{a}\vec{n}_{ab}$. The new normals $\vec{n}'_{ab}$ satisfy \Ref{vecgeo}, and the results derived above immediately apply. We only have to rotate the final expression back to the original $\vec n_{ab}$, and in doing this we pick up a phase
from the component of the rotation $R_a'$ along the direction $\vec n_{ab}$. This can be computed to be
\be\label{extraphase}
\d_{ab} = \arctan \left(\f{\vec v_a\cdot\hat z}{(1+\vec n_{ab} \cdot\hat z)\tan\psi_a/2 + \vec n_{ab}\times \vec v_a\cdot\hat z} \right),
\ee
so that (\ref{LO}, \ref{LO2}) are corrected by an extra global phase $\sum_{ab}j_{ab}\d_{ab}$.
Again, it can be reabsorbed re-phasing the coherent states, as explained in \cite{BarrettSU2}. Hence, it is always possible in this way to get real leading order behaviours, non-oscillating like \Ref{LO} for non-shape-matched vector geometries, and a cosine like \Ref{LO2} without the global phase for Regge geometries. Our numerics below will show that this reality extends well beyond the leading order.

Finally, if the normals are such that there is no solution to the critical point equations \Ref{CP1}, then there are no saddle points and the amplitude is exponentially suppressed. 

The results have been derived for the definition \Ref{Av} of the coherent amplitude, but everything extends trivially to the holomorphic version \Ref{Aholo} used in \cite{BarrettSU2}. The difference between the two is only a global phase, whose irrelevance we have commented upon and that can also be reabsorbed in the definition of the coherent states to obtain a real result.
Up to the phase difference, both definitions contain a parity transformation in the map between the two classical vectors and the two kets associated with a link, and this is what simplifies the saddle point analysis. Had we worked with bras $\bra{\vec n_{ab}}$ without any parity transformation, the trivial saddle point would have shifted from the identity to the parity group element $\eps$, thus unnecessarily complicating the analysis.

We have thus recovered the results of \cite{BarrettSU2}. 
The global phase of the leading order is gauge-covariant, and can be arbitrarily changed re-phasing the coherent states. Only when two distinct saddle points exist, the relative phase is unambiguous and has a  meaning intrinsic to the $15j$.
The frequency of these oscillations is given by the Regge (boundary) action for a flat 4-simplex, and this is the most important aspect of the asymptotic behaviour. In the next Section, we present numerical calculations supporting these results, and providing additional information on their accuracy and next-to-leading order behaviours.

\section{Numerical results}\label{NumEv}
To proceed with a numerical evaluation of \Ref{Av}, there are two possible approaches that one can take: either performing the integrals in the first equality with MonteCarlo techniques; or perform an exact evaluation of the $15j$ symbols summed over the coefficients of the coherent states, as 
in the second equality. 

The first approach is a priori faster, however adaptive methods are required to deal with the oscillations, and the convergence can be slow already with the simpler invariants used in the Barrett-Crane model \cite{Christensen:2001eu,IoDan,IoDan2}. Furthermore, our main motivation for numerical studies of the coherent $15j$ asymptotics was as a warm-up exercise for the Lorentzian EPRL model \cite{EPRL}, and in that case the MonteCarlo approach suffers further from the non-compactness of the group integrals. This difficulty can be reduced using the factorisation property of $\SL(2,\C)$ Clebsch-Gordan coefficients \cite{Boosting}, but the technique requires to handle summations over SU(2) $nj$ symbols. With these considerations in mind, we used for this paper the exact summation approach. We postpone a study of MonteCarlo techniques for such vertex amplitudes to future work.

\begin{figure}[ht]
    \centering
        \includegraphics[width=7cm]{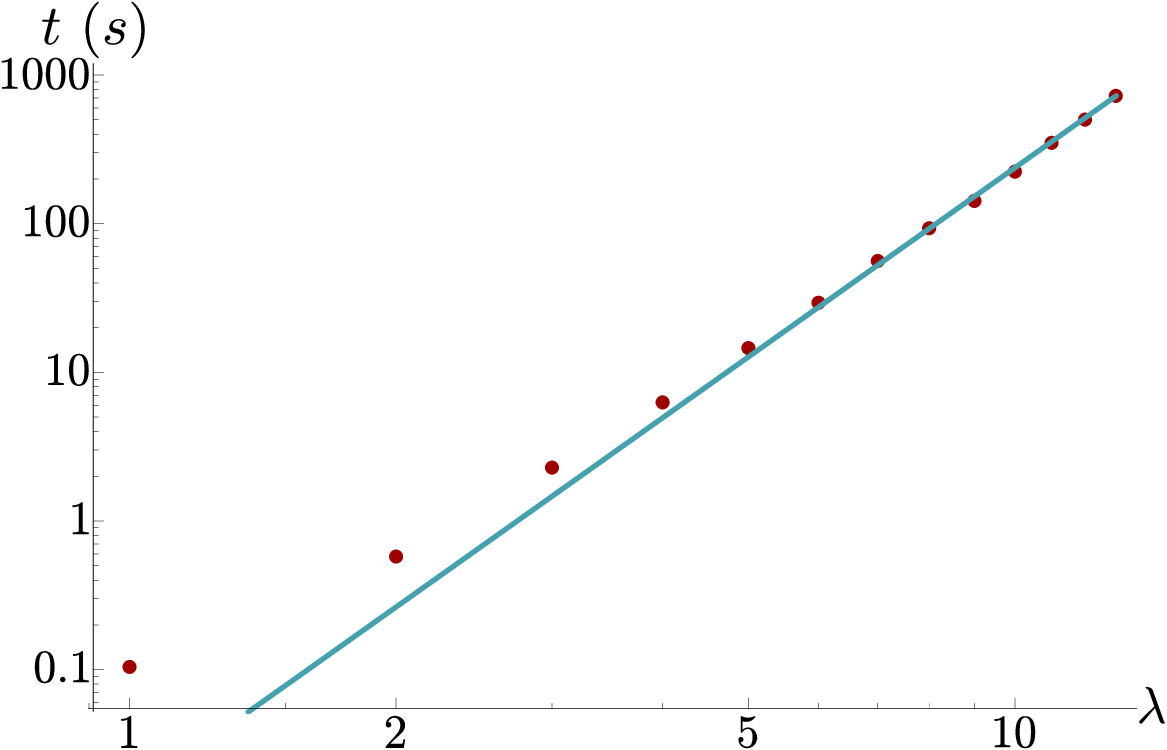}
    \caption{        \label{FigTimings} 
    \small{\emph{Evaluation times for the equilateral configuration of Fig. \ref{FigLO}. The log-log plot shows a power law increase in the needed time, with power law that can be fitted by $0.01 \lambda ^{4.36}$. The extrapolated time for $\lambda = 25$ which is the last point of Figure \ref{FigLO} is of about three hours and a half. Note that this estimation is done with $3jm$'s, $6j$'s and $9j$'s already calculated and pre-loaded into RAM. }} }
\end{figure}

\begin{figure}[ht]
    \centering
    \begin{subfigure}[t]{0.49\textwidth}
        \includegraphics[width=7.5cm]{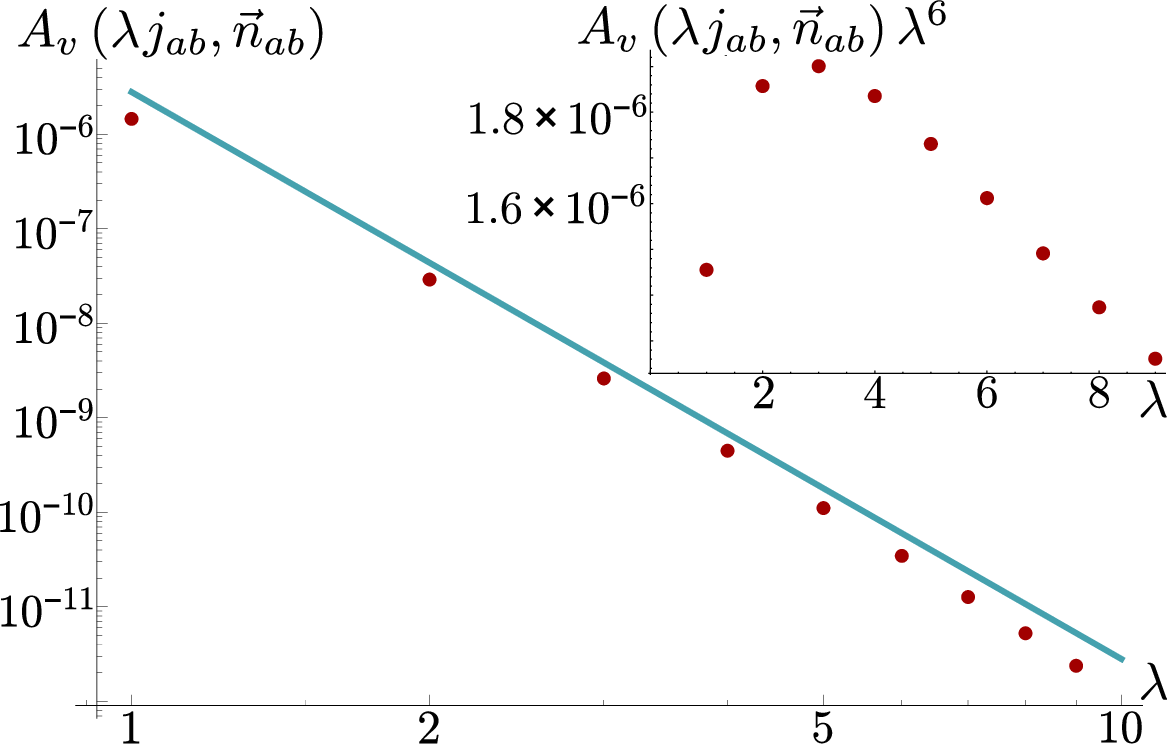}
    \end{subfigure}
    \begin{subfigure}[t]{0.49\textwidth}
        \includegraphics[width=7.5cm]{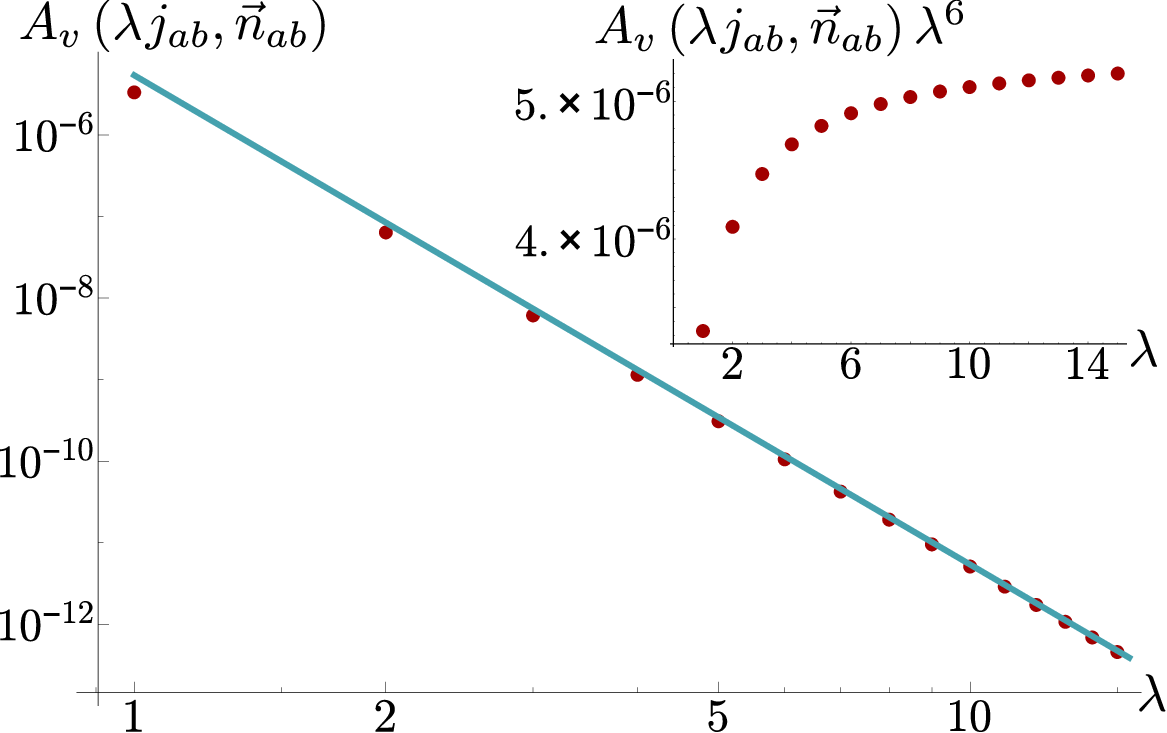}
    \end{subfigure}
    \caption{\small{Left panel: \emph{An example of saddle-less configuration showing exponential decay of the amplitude. Log-log plot in the main picture, with data points in red and a $\l^{-6}$ plotted line for comparison. A linear plot is also shown in the upper picture. The boundary data are those of a Lorentzian 4-simplex, which satisfy closure but not the orientation condition \Ref{CP1}. } Right panel: \emph{A vector geometry configuration with a single critical point, showing a power law decay $\l^{-6}$. Log-log plot in the main picture, with data points in red and the analytic result \Ref{LO} and $\propto \l^{-6}$ plotted for comparison. A linear plot is also shown in the upper picture. The boundary data are $j_{ab}=2$ except $j_{23}=j_{45}=3$, and angles $\varphi_{53,42}=\frac{\pi}{4}$, $\varphi_{34,5}=\varphi_{25,4}=\varphi_{25,3}=\varphi_{34,2}=\frac{\pi}{8}$ in the notation developed in Section \ref{SecVecGeo}.}} \label{deontologia} }
\end{figure}

\begin{figure}[ht]
    \centering
    \begin{subfigure}[b]{0.49\textwidth}
        \includegraphics[width=7.5cm]{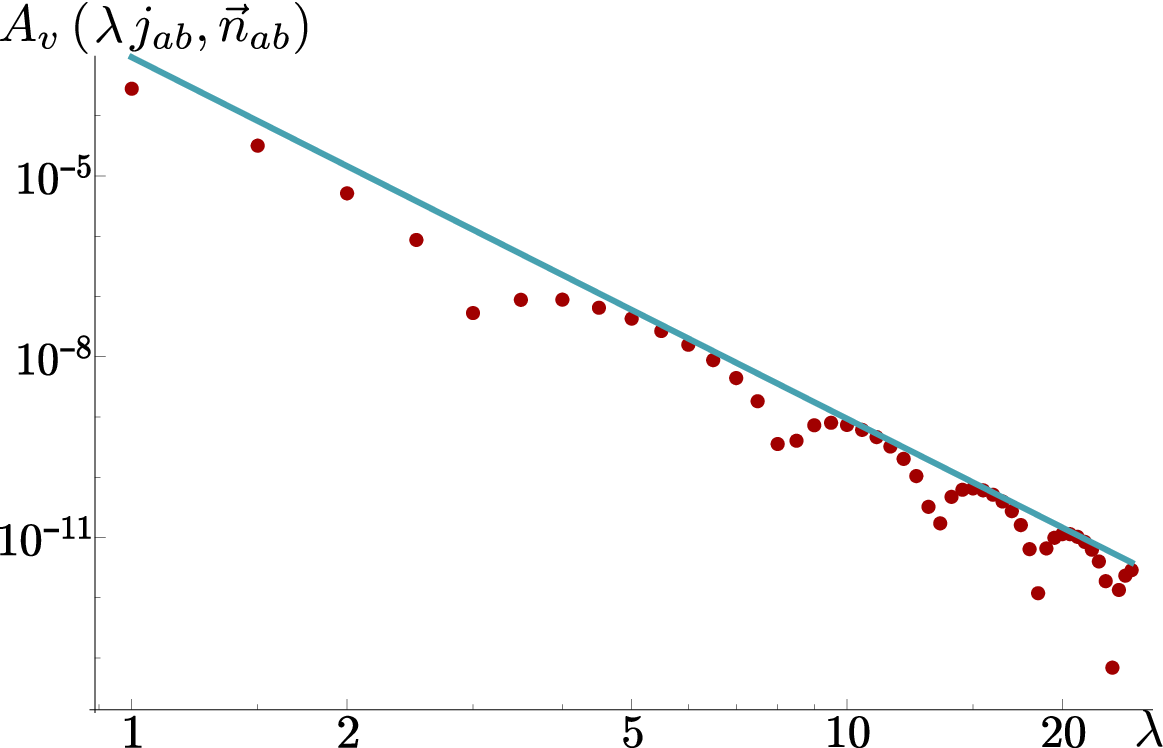}
    \end{subfigure}
    \begin{subfigure}[b]{0.49\textwidth}
        \includegraphics[width=7.5cm]{_images/SU2DataCos.eps}
        \label{SU2DataCos}
    \end{subfigure}

\caption{\label{FigLO} { \small{\emph{Numerical data points versus the analytic leading order \Ref{LO2} for equilateral Regge data, log-log (left panel) and linear (right panel) plots.} }} }

\end{figure}

\begin{figure}[ht]
    \centering
    \begin{subfigure}[b]{0.49\textwidth}
        \includegraphics[width=7.5cm]{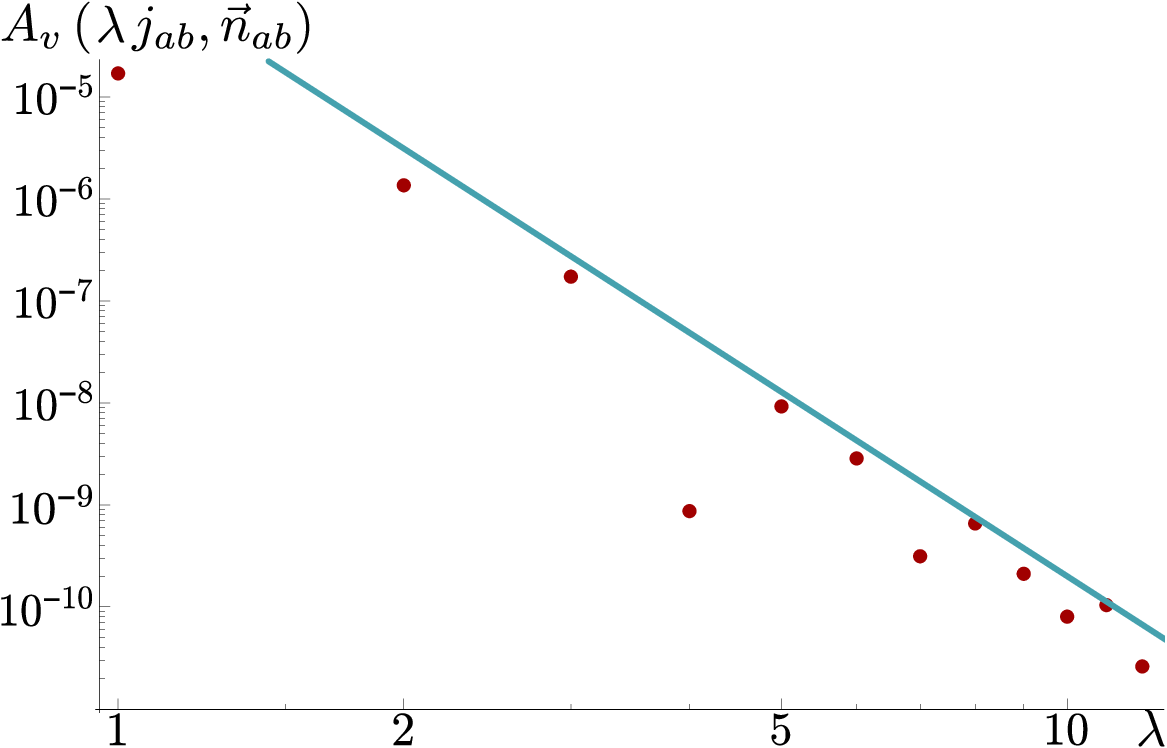}
        \label{SU2DataPowerLaw2}
    \end{subfigure}
    \begin{subfigure}[b]{0.49\textwidth}
        \includegraphics[width=7.5cm]{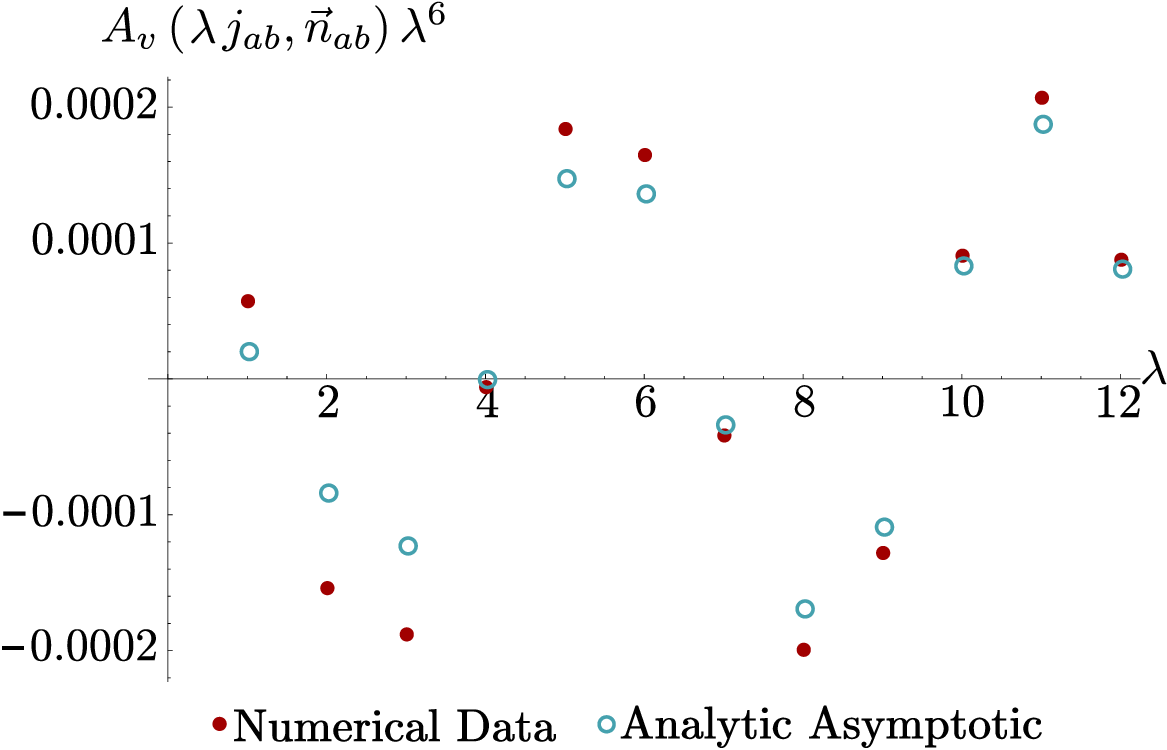}
    \end{subfigure}
    \caption{\small{\emph{Numerical data points versus the analytic leading order \Ref{LO2} for isosceles Regge data, log-log (left panel) and linear (right panel) plots. We provide less data points than for the equilateral configuration of Fig.~\ref{FigLO} because of the slower evaluation times (caused by the ratio 2-to-1 between the spins) and of the absence of half-integer spins, not allowed for this configuration.}} 
    \label{FigPlot2}}
\end{figure}

With this approach, the main difficulty comes from the power-law increase of  $15j$ symbols required to compute the sum over intertwiner labels $i_a$ in \Ref{Av}. For instance, for an equilateral configuration with all spins equal to $j$, we need $(2j+1)^5$ different $15j$ symbols for each data point. It is thus adamant to be as efficient as possible in the evaluation of each symbol. 
If we started from the definition \Ref{15j} in terms of Wigner's $3jm$ symbols, evaluating the asymptotics would be an impossible task. To give an idea, with a 2,4 GHz CPU running Wolfram's Mathematica, the best timing we could reach\footnote{To optimise the calculation, we stored all the needed $4jm$ symbols in the RAM, and reduced the summation over magnetic indices implementing in the algorithm the Clebsch-Gordan selection rules.} is order $10^{-2}$ seconds to compute one $\{15j\}$ for $j_{ab}=i_a=1$, but this quickly grows to about 9 seconds for $j_{ab}=i_a=5$. At those still very small spins, we would already need approximately 17 days to sum $\approx 10^5$ of them.
The trick to avoid this obstacle is the observation that we do not need to evaluate the irreducible $15j$ symbols \Ref{15j}. In fact, the sum over coherent intertwiners is invariant under change of basis. We can thus choose a basis of reducible $15j$ symbols, whose evaluation is much faster. In particular, we took as basis the following reducible symbol (related to \Ref{15j} by two recoupling moves on $i_2$ and $i_5$), and which factorises into two $6j$'s and one $9j$ symbol:
\begin{equation}
\begin{split}\label{15jr}
\raisebox{-16mm}{\includegraphics[height=3.5cm]{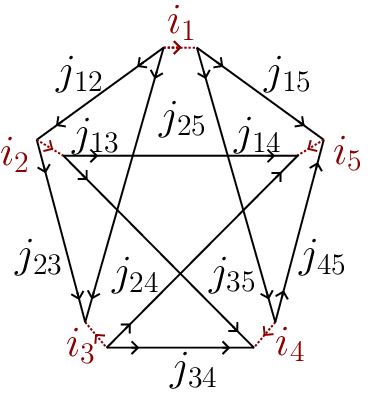}} =  
(-1)^\chi \ 
\raisebox{-14mm}{\includegraphics[height=3cm]{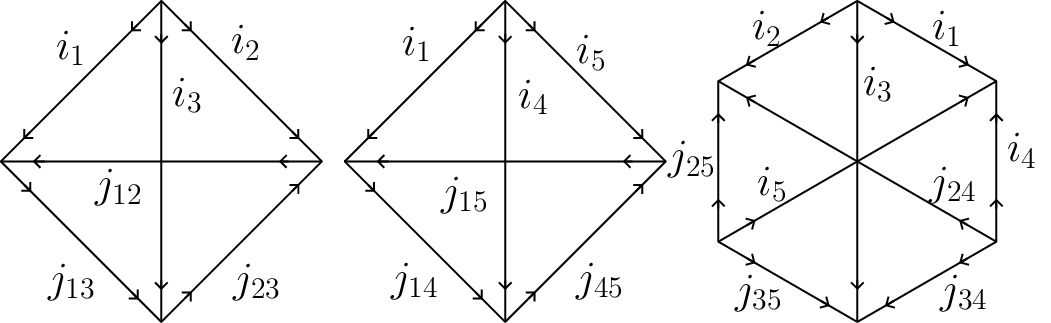}}
\end{split}
\end{equation}
with the phase given by $\chi:={2 (j_{13} + j_{15}) + 2i_1 - (i_2 + i_3 + i_4 + i_5)}$.
Then, the evaluation of $6j$'s and $9j$'s can be performed in virtually instantaneous machine time using the $C-$based softwares \texttt{wigxjpf} and \texttt{fastwigxj} developed in \cite{Johansson:2015cca}, and to which we refer for more precise evaluation times and estimates of the numerical errors.\footnote{Wigner's functions are evaluated with machine precision using Mathematica. Summations are performed with the command Compensate to achieve again machine precision. Hence, the only meaningful numerical errors of our results are those that can come from  \cite{Johansson:2015cca}.}
Adapting them to properly interact with out Mathematica code, each \Ref{15jr} takes about $10^{-6}$ seconds. This enormous improvement makes the approach feasible.
To complete the the computation of the amplitude, we need to weight the reducible $\{15j\}$ symbols with the coherent states \Ref{coeffCS} and sum over the intertwiners.  The number of $15j$ so required increases as a power law, and so does the evaluation time. 
Fig. \ref{FigTimings} provides a measure of the time required for each data point in the equilateral configuration of area $j$. 
To evaluate the amplitude numerically, we used \Ref{Av} with Perelomov's coherent states. We present in this Section data plots covering all three asymptotic behaviours: saddle-less generic boundary data, non-shape-matched vector geometries, and Regge data. For all cases studied we confirmed the estimates of \cite{BarrettSU2} to very good accuracy, and found that the leading order global phase is exact to all orders, within numerical precision at least. Examples of the exponential suppression for generic boundary data, and non-oscillating power-law \Ref{LO} for vector geometries, are shown in Fig.~\ref{deontologia}. For Regge data, we present two examples, the equilateral configuration in Fig.~\ref{FigLO}, and the isosceles configuration of \Ref{isosceles} (an equilateral tetrahedron with spins equal $2\l$ and four isosceles tetrahedra of spins $2\l$ and $\l$) in Fig.~\ref{FigPlot2}. 

\subsection{Regge data, leading order}\label{SecLO}

Let us first point out a technical detail on our choice of data for the numerics. When choosing areas and normals for Regge data, it is convenient to fix the relative orientations of the tetrahedra in order to make the normals pairwise antiparallel as in \Ref{vecgeo}. This avoids the presence of the extra global phase \Ref{extraphase} in the asymptotics. For a different configuration of normals corresponding to Regge data one has to determine anyways a set of rotations that makes the normals pairwise antiparallel in order to evaluate \Ref{extraphase} and get the right global phase of the asymptotic.\footnote{One can of course bypass this point looking only at the absolute value of the data and the amplitude, however we wanted to test also the global phase of the formula, especially in the light of studying the next-to-leading order.} %
For instance for the equilateral configuration, the simplest choice of data would be to take five identical copies of the same configuration of normals. The asymptotic will however depend explicitly via the global phase on the rotations making the normals pairwise antiparallel. It is then simpler to start directly from the pairwise antiparallel configuration, and compare the numerical data to \Ref{LO2}. This is what we did for our numerical studies.
The configuration with all normals pairwise antiparallel can be represented by a 3d object that we refer to as twisted spike, whose construction and properties will be described in Section 4.1 below. The explicit configuration of normals used for the equilateral configuration is reported in Appendix~\ref{AppA}. 

For Regge data, our numeric analysis tests three properties of the asymptotic formula \Ref{LO2}: the $\l^{-6}$ slope, the oscillations with frequency given by the Regge action, the offset given by the phase of the Hessian. 
The first two were computed in \cite{BarrettSU2}, the last one we provided above on a case-by-case basis thanks to \Ref{daje}. All three results are confirmed to very good accuracy, see for examples the data for the equilateral and isosceles configurations in Figs.~\ref{FigLO} and \ref{FigPlot2}. 
Notice that in these plots the numerical data have been divided by the leading-order global phase $e^{i \lambda S_R }$, to have a real leading order. This division could leave a priori an $O(\l^{-7})$ imaginary part in the data, but the results of numerics are \emph{exactly} real, at least to machine accuracy at $O(10^{-22})$. This suggests that the leading order phase may be exact to all orders, as anticipated at the end of Section 2.5, where we also provided a possible way to prove it.

\begin{figure}[ht]
    \centering        \includegraphics[width=7.5cm]{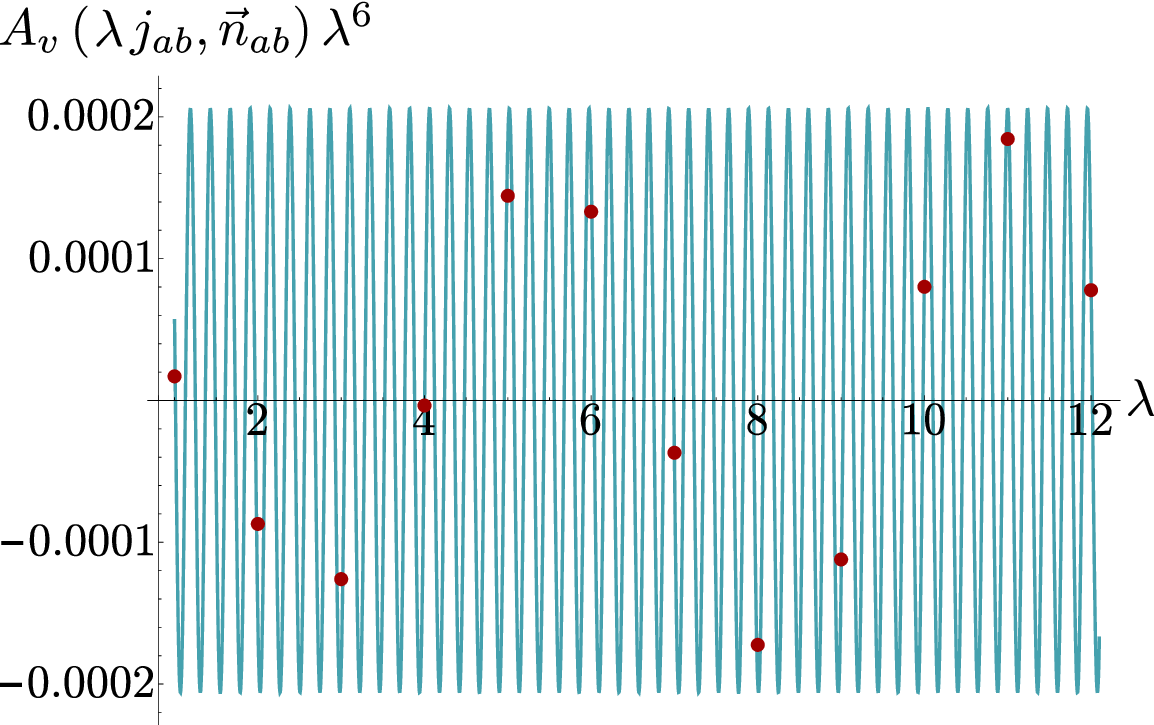}
       \caption{\label{Figreal}  \small{ \emph{The rescaled data (for the equilateral configuration) and asymptotic formula with real arguments, showing that the real frequency is much higher that what could be naively inferred from looking only at the integer values. A similar situation occurs in the asymptotics of the $6j$ symbol.} } }
\end{figure}

To improve the visibility of the data, we have plotted in the figures the asymptotic formula only for integer arguments. The real frequency of the asymptotic formula is however much higher that what a naive interpolation may suggest, as can be seen in Fig.~\ref{Figreal}, a familiar feature from the $6j$ case.

To estimate the accuracy of the asymptotic formula, we looked at the relative error, defined as the difference between the exact evaluation and the leading order, normalised by the exact evaluation. See left panel of Fig.~\ref{FigNLO}. This shows that we already have agreement to a few per cent level at spins of order $10$.
This accuracy is comparable to that of the Ponzano-Regge formula for the $6j$.\footnote{The situation will be different for the Lorentzian EPRL model based on $\SL(2,\C)$, where for Lorentzian boundary data much higher spins will be needed \cite{noiLor}.}

\subsection{Regge data, higher orders}

\begin{figure}[ht]
    \centering
        \includegraphics[width=7.5cm]{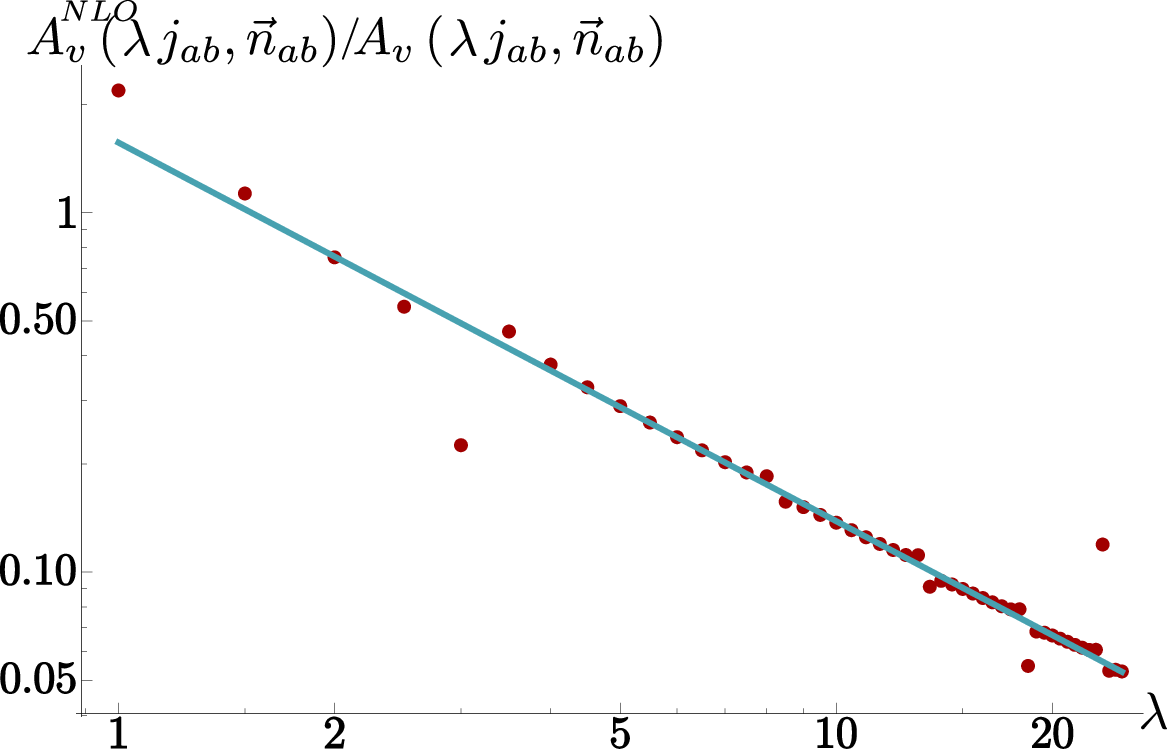}
\hspace{1cm}                \includegraphics[width=7.5cm]{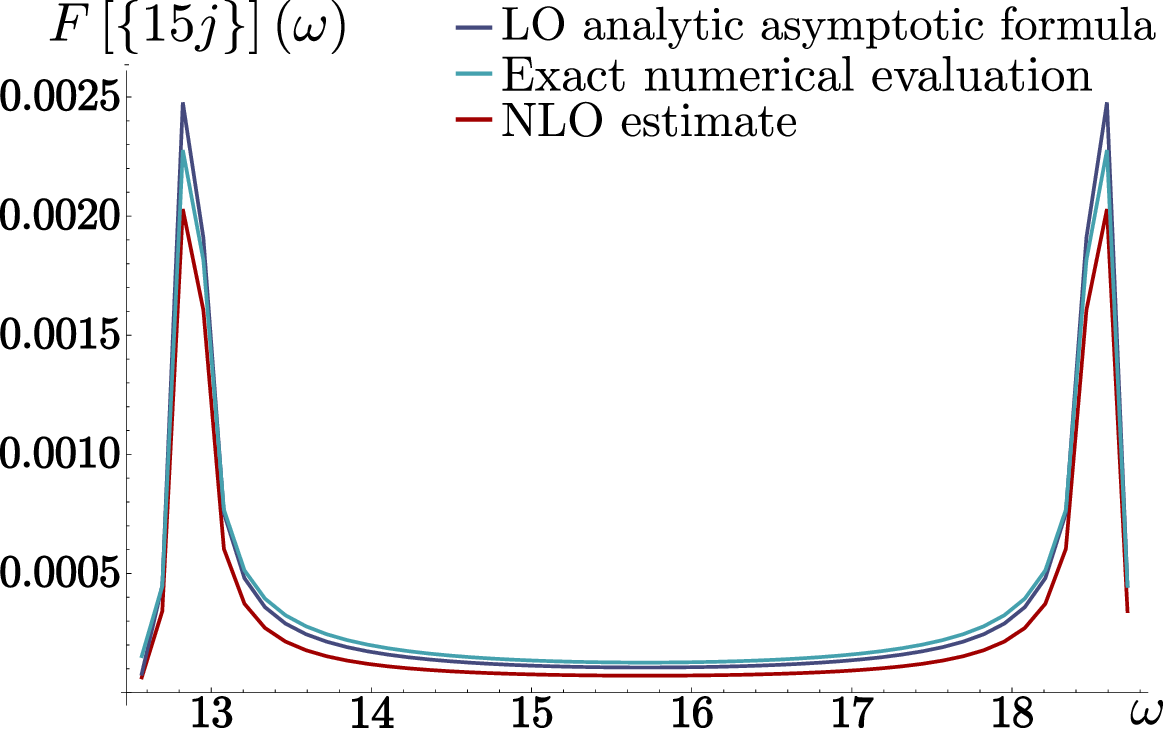}
       \caption{\label{FigNLO}  \small{Left panel: \emph{Log-log plot of the ratio between of the next-to-leading order and the exact evaluation of the amplitude (red dots) as a function of the scale $\lambda$, compared to the power law $\l^{-1}$ (blue line) predicted by the asymptotic formula.} Righ panel: \emph{Discrete Fourier analysis of the analytic leading order formula, and the exact numerical evaluation, and of the estimated next-to-leading order (NLO) of the left panel, showing they all have the same frequency. The presence of two symmetric peaks is a trivial consequence of the properties of the discrete Fourier transform of a real signal. For the analytic formula this is to be expected, and the plot shows that also the exact evaluation and estimation of the NLO are exactly real to within machine accuracy. The three plots have different vertical shifts because the signal is not normalised. } } }
\end{figure}

The validity of the saddle point approximation \Ref{LO2} also makes precise predictions about the higher order terms: they are organised in increasing inverse powers of $\l$, and share the same frequency of oscillations (but a priori different global phases and phase offsets). To provide some numerical testing of these properties, we looked at the next-to-leading order by subtracting from the numerical data the corresponding value obtained from the asymptotic formula. The result for the equilateral configuration is shown in Fig.~\ref{FigNLO}. The left panel shows agreement with the $O(\l^{-7})$ scaling. To confirm the same frequency of oscillations as the leading order, we performed a discrete Fourier analysis of the data, shown in the right panel.

The numerics confirm the good validity of the saddle point expansion for the Regge data considered. If we extrapolate this expansion to all orders, we expect an asymptotic series with the following structure:
\be\nn
A_v(j_{ab},\vec n_{ab}) = \sum_{n=0}^\infty \frac{1}{\lambda^{6+n}}\, \left(c_n(j_{ab},\vec n_{ab}) e^{i\l S_R(j_{ab},\vec n_{ab})+i\Phi_n(j_{ab},\vec n_{ab})} +d_n(j_{ab},\vec n_{ab})e^{-i\l S_R(j_{ab},\vec n_{ab})+i\Psi_n(j_{ab},\vec n_{ab})} \right),
\ee
with $c_n, d_n,\Phi_n,\Psi_n$ real functions. From the explicit calculations of Section \ref{SecCritical} we know that $c_0=d_0$, $\Phi_0=\Psi_0$, which leads to the cosine form \Ref{LO2} of the leading order. It is possible that similar relations hold to all orders, thus making the series a sum of cosine and sine terms. For instance, our numerical data in Fig.~\ref{FigNLO} show the reality of the next-to-leading order, and two symmetrical peaks in the Fourier transform. This indicates that $c_1=d_1$ and $\Phi_1=\Psi_1$, hence
\be
A^{NLO}_v(j_{ab},\vec n_{ab}) = \f{2 c_1(j_{ab},\vec n_{ab})}{\l^7} \cos \Big( \l S_R(j_{ab},\vec n_{ab})+\Phi_1(j_{ab},\vec n_{ab}) \Big).
\ee
This extrapolation can be compared with the expansion to all orders for the isosceles $6j$ symbol explicitly computed in \cite{Bonzom:2008xd} (see also \cite{Kaminski:2013gaa}), which shares the same structure. 

The fact that the same Regge action appears at all orders has implications for quantum gravity models based on these amplitudes:  the large spin limit is interpreted as a semiclassical $\hbar G\mapsto 0$ limit, and one could ask whether the saddle point approximation is related to an expansion in higher order curvature invariants of the metric. This would actually be hard to achieve in the framework of Regge calculus, a discretisation that cannot distinguish for instance the square of the Ricci tensor from the square of the Ricci scalar, prior to taking the continuum limit. In any case, the saddle point expansion shows that the quantum corrections do not add extra curvature invariants to the action, but rather contribute to the measure of the path integral. This can be understood if we recall that the quantum amplitude arises from a first order path integral. The saddle point approximation is thus computing the contribution to the measure arising from integrating over the connection degrees of freedom.\footnote{See also \cite{Dittrich:2014rha} for additional discussions on the BF versus Regge measure term in the path integral.}

\section{The twisted spike and vector geometries}
In this Section we study a few more properties of the geometries described by the boundary data, providing a 3d picture for those corresponding to a 4-simplex, and a characterisation of vector geometries as a special class of twisted geometries.

\subsection{The twisted spike}\label{SecGeo}
Barrett's reconstruction theorem, or the related procedure described above, shows how to reconstruct a 4-simplex from a shape-matching configuration of areas and closing normals. But now consider the opposite question: given a 4-simplex described by its lengths (or equivalently its areas), how do we derive a compatible set of 3d normals? This `deconstruction' was necessary to us to provide explicit boundary data for the numerical tests of the previous Section, but it is also useful to gain more insight on the geometry of the boundary data and leads us to a contact between the dynamics and the canonical theory.

\begin{figure}[ht]   
 \centering
    \begin{subfigure}[t]{0.45\textwidth}
        \includegraphics[width=0.9\textwidth]{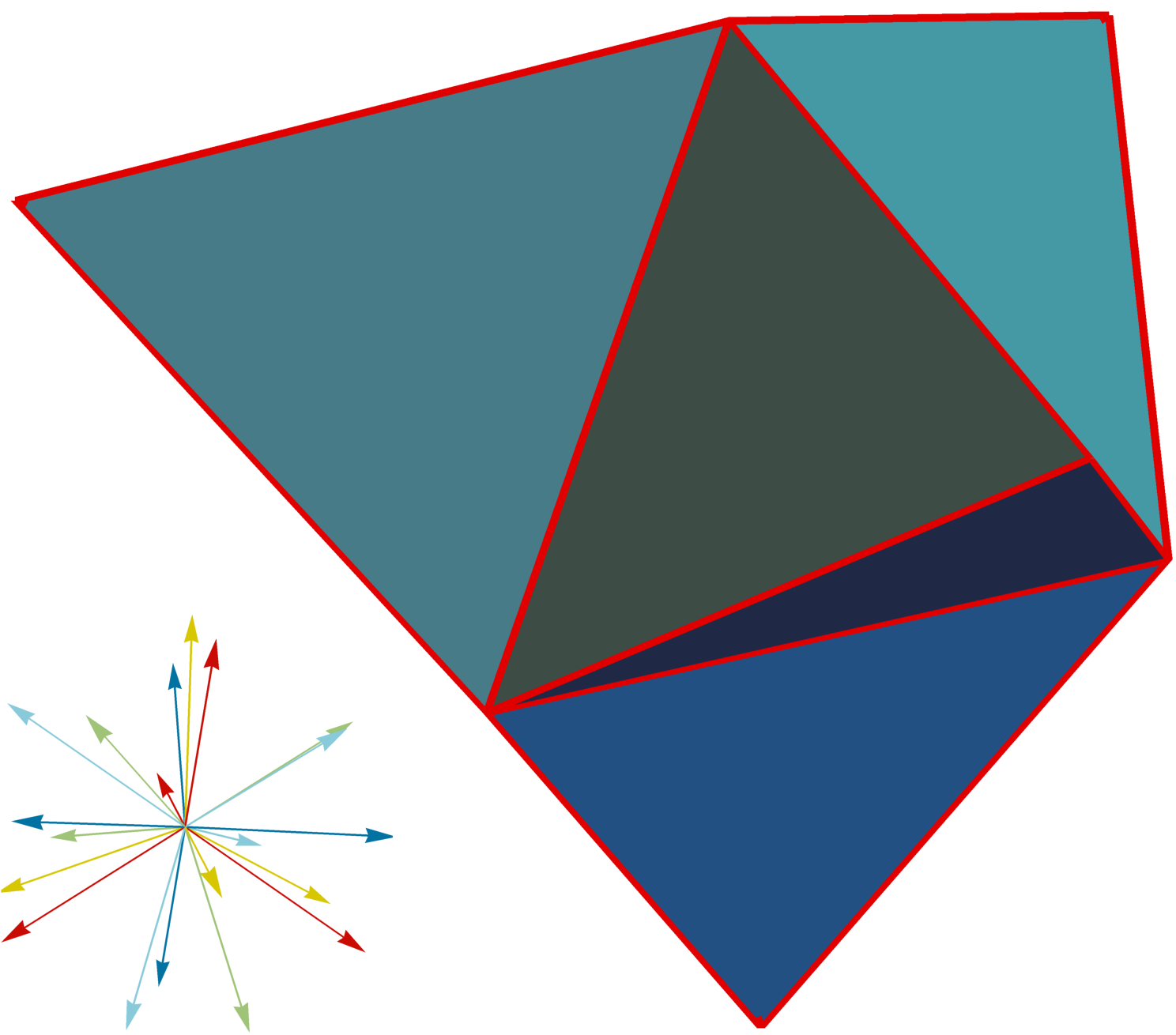}
    \end{subfigure}
    \begin{subfigure}[t]{0.45\textwidth}
        \includegraphics[width=.95\textwidth]{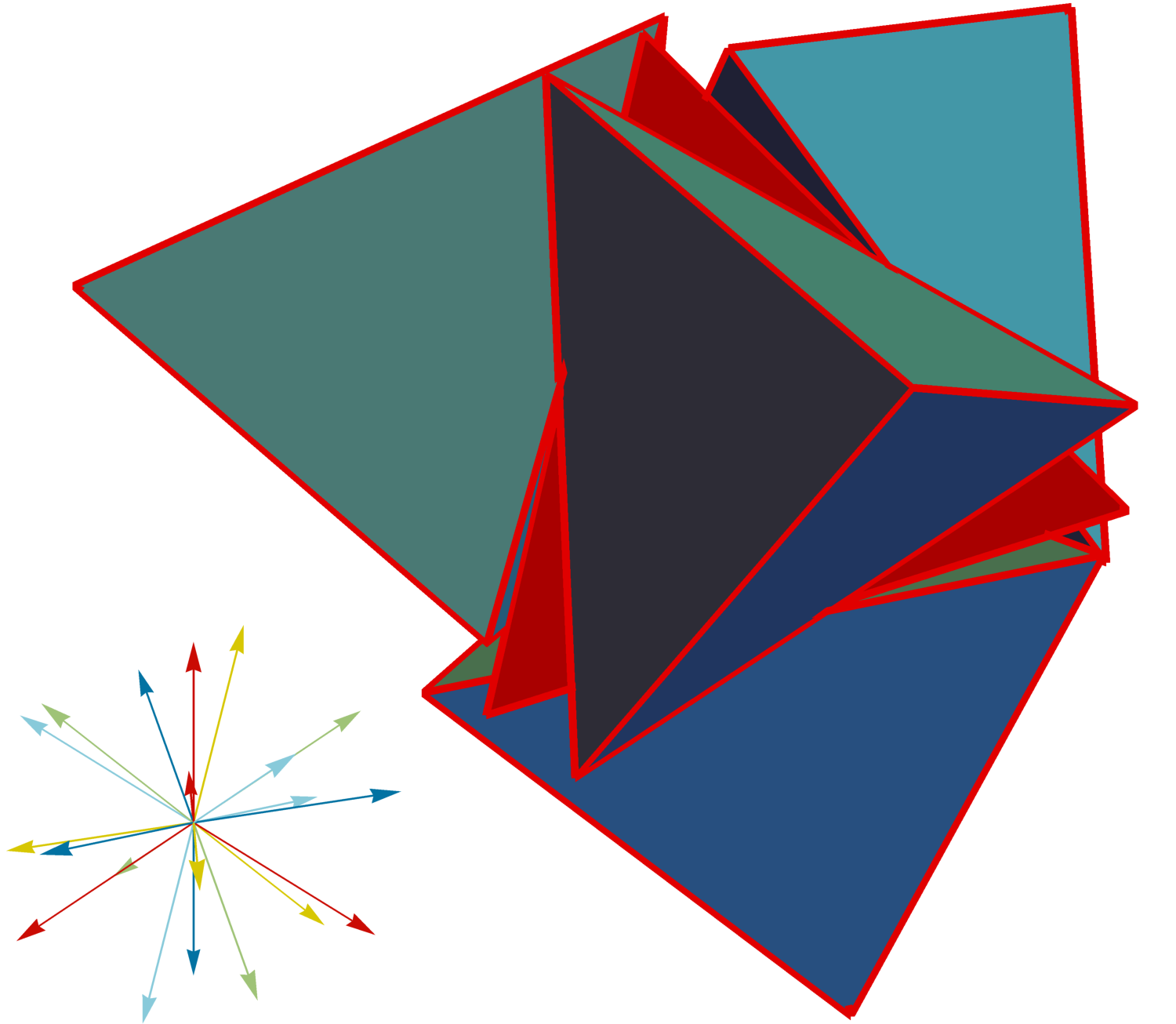}
    \end{subfigure}
\caption{\label{FigSpikes} {\small{Left panel: \emph{the `spike', an Euclidean 4-simplex shown in $\R^3$ by means of rotating four tetrahedra in the same 3d space of a reference tetrahedron, here the central one (the four outmost vertices would be identified in $\R^4$). The associated normals are shown in the bottom left picture, equal colour for those belonging to the same tetrahedron.} Right panel: \emph{the `twisted spike', a rotation in the plane of each shared triangle by an equal amount to the dihedral angle makes the triangle normals all opposite to one another.} }} }
\end{figure}

Consider an Euclidean 4-simplex, and denote  $N_a$ the 4d unit vectors normal to the tetrahedron $a$, so that the external dihedral angle between two tetrahedra is $\theta_{ab}=\arccos \left(N_a\cdot N_b\right)$. To project it to 3d, let us pick the tetrahedron $1$ as reference, and rotate in $\R^4$ the remaining tetrahedra as to align their normals to $N_1$. Each of these rotations preserve the triangle shared between the tetrahedron $a$ and $1$, and the result is a collection of five tetrahedra all glued to the triangles of the tetrahedron $1$ that is contained in the three dimensional hyperplane orthogonal to $N_1$. We refer to this 3d geometric object as the \emph{spike}, see left panel of Fig. \ref{FigSpikes}. 
The five tetrahedra share now a common $\R^3$ frame and we can compute the outgoing 3d normals to the triangles. By construction, the normals to the triangles shared with the tetrahedron $1$ are opposite to one another, $\vec n_{1a}=-\vec n_{a1}$, but not the others. Consider then a rotation of each tetrahedron around the axis of the normals shared with 1, of an angle given by the 4d dihedral angle $\th_{1a}$. 
As we know from the saddle point analysis in Section \ref{SecCritical}, the result of this rotation is to make all normals pairwise anti-parallel, as in \Ref{vecgeo}. We refer to this geometric object as the \emph{twisted spike}, shown in the right panel of Fig. \ref{FigSpikes}.  

From this perspective, it may look surprising that precisely a rotation of the dihedral angle has the property of making the normals all opposite;
but this is fairly easy to show. Consider the angle between two edge vectors in the common $\R^3$ frame of the spike: it is straightforward to show that when \Ref{vecgeo} holds this angle coincides with the 4d dihedral angle obtained from the spherical cosine laws \Ref{thtophi}, 
\be\label{Pietro}
\frac{\left(\vec{n}_{1b}\times\vec{n}_{1a}\right)\cdot \left(\vec{n}_{a1}\times\vec{n}_{ab}\right)}{\left\Vert \vec{n}_{1b}\times\vec{n}_{1a}\right\Vert \left\Vert \vec{n}_{a1}\times\vec{n}_{ab}\right\Vert } \stackrel{\Ref{vecgeo}}{\equiv} \cos\th_{1a}^{(b)}(\vphi).
\ee

Here we used the common $\R^3$ frame, but the property that the twist angle between edge vectors coincides with the 4d dihedral angle at the saddle point is gauge-invariant. To see that, we define the gauge invariant twist angle as in \cite{DittrichRyan,IoMiklos}, and using this time \Ref{CP1} we have
\be\label{defxi}
\cos\xi_{1a}^{(b)}:=\frac{\left(\vec{n}_{1b}\times\vec{n}_{1a}\right)\cdot R_1R_a^{-1}\left(\vec{n}_{a1}\times\vec{n}_{ab}\right)}{\left\Vert \vec{n}_{1b}\times\vec{n}_{1a}\right\Vert \left\Vert \vec{n}_{a1}\times\vec{n}_{ab}\right\Vert } \stackrel{\Ref{CP1}}{\equiv} \cos\th_{1a}^{(b)}(\vphi).
\ee
This 3d twisting thus records a 4d rotation preserving a triangle in terms of a 2d rotation in the plane of that triangle. 
For the applications of the asymptotic results to loop quantum gravity, it is important to stress that this twisting is the discrete counterpart of the use of self-dual Ashtekar variables in general relativity (here for Euclidean signature). In particular, \Ref{totti} encodes at the discrete level the secondary simplicity constraints guaranteeing the torsionlessness of the connection, as discussed in \cite{DittrichRyan,IoFabio,IoMiklos}.\footnote{The construction can also be extended to real Ashtekar-Barbero variables, in which case the exact embedding of SU(2) in SO(4) or $\SL(2,\C)$ depends on the Immirzi parameter $\g$ \cite{DittrichRyan,IoFabio,IoMiklos}. For the Lorentzian case the 4d dihedral angle is a boost $\Xi_{1a}$, and the shape-matched solution of the discrete secondary simplicity constraints gives 
\[
\xi_{1a}^{(b)} = \f1\g \Xi_{1a}, \qquad N_1\cdot N_a = -\cosh\Xi_{1a}.
\]
}

For completeness and to make contact with the procedure of \cite{BarrettSU2}, let us briefly touch upon the bivector description of the geometry.
Given an Euclidean 4-simplex, from the areas and 4d normals to the tetrahedra we can define a set of 20 simple bivectors as follows, 
\[
B_{ab}=j_{ab}\star\frac{N_{a}\wedge N_{b}}{\left\Vert N_{a}\wedge N_{b}\right\Vert }, \qquad \qquad B_{ab}^2 = j_{ab}^2,
\]
or in components $B_{ab}^{IJ}=(j_{ab}/{2\sin\theta_{ab}})\epsilon_{\phantom{IJ}KL}^{IJ}N_{a}^{K}N_{b}^{L}$. Since we derived them from a geometric 4-simplex, they also satisfy cross-simplicity $\eps_{IJKL} B^{IJ}_{ab} B^{KL}_{ac}$ and non-degeneracy, as well as the orientation and closure conditions
\be\label{bivectors}
B_{ab}=-B_{ba}, \qquad
\sum_{b\neq a}B_{ab}=0 \quad \forall a.
\ee
These conditions guarantee that the self-dual and anti-self-dual components of the bivectors coincide up to an SO(3) rotation. We can thus write
\be
B_{ab} = j_{ab} (\vec n_{ab}, R_a \vec n_{ab}),
\ee
for some unit vectors $\vec n_{ab}$ and SO(3) rotations $R_a$. Then \Ref{bivectors} immediately imply the closure and orientation equations \Ref{CP} for either self-dual or anti-self-dual parts, so the geometric 4-simplex defines a saddle point.

Using bivectors, we can also get the mapping between the twist angle and the 4d dihedral angle as defined by the scalar product of 4-normals, without passing for the spherical cosine law. To that end, let us choose a 4d normal, $N_1$, to define a unique self-dual projector for all bivectors:
$N_{1I} P^{IJ}_{KL}:=N_{1I}(\delta^{IJ}_{KL}+\frac{1}{2}\eps^{IJ}{}_{KL})/2.$ 
Then, taking the 3d normal as the self-dual parts of the bivectors,  
\be\label{manovella}
j_{ab}n_{ab}^J:=N_{1I} P^{IJ}_{KL} B_{ab}^{KL} = j_{ab}\frac{1}{4\sin\theta_{ab}}N_{1I}\left(\delta_{KL}^{IJ}+ \f12\eps^{IJ}{}_{KL}\right)N_{a}^{K}N_{b}^{L}
\quad \stackrel{\perp N_1}{\longrightarrow}  \quad j_{ab}\vec{n}_{ab}.
\ee
Using $\eps_{ijk} = N_1^I \eps_{Iijk}$ we can lift \Ref{Pietro} to a covariant formula, and after some lengthy but straightforward algebra, we get
\be
\frac{\left(\vec{n}_{1b}\times\vec{n}_{1a}\right)\cdot \left(\vec{n}_{a1}\times\vec{n}_{ab}\right)}{\left\Vert \vec{n}_{1b}\times\vec{n}_{1a}\right\Vert \left\Vert \vec{n}_{a1}\times\vec{n}_{ab}\right\Vert } \stackrel{\Ref{manovella}}{\equiv} N_{1}\cdot N_{a}.
\ee
This proves that \Ref{bivectors} implies that $(i)$ the twist angle is the same for every edge of the chosen triangle, and $(ii)$, it coincides with the 4d dihedral angle.
The exactly same formula can be derived choosing the anti-self-dual projection, and using the expression \Ref{defxi},
\be\label{totti}
\cos\xi_{1a}^{(b)}= N_{1}\cdot N_{a}.
\ee

\subsection{Vector geometries}\label{SecVecGeo}
Since vector geometries are a subset of twisted geometries, they can be interpreted as a collection of mismatched tetrahedra. One can then ask whether it is possible to span the space of gauge-invariant vector geometries using 5 independent shape variables. The answer turns out to be negative, because the shape mismatch is in general quite complicated. The best parametrisations we could find are either 6 shape parameters subject to a non-linear constraint, or 4 shape parameters plus a fifth non-gauge invariant parameter.

For the first parametrisation, we start by fixing the $10$ areas and the $2$ shape parameters of three tetrahedra, and try to build a vector geometry configuration. Using the freedom to  independently rotate the three tetrahedra, we can always arrange the normals on the shared triangles to be opposite to one another. Consider next the fourth and fifth tetrahedra. We have not specified yet any of their normals, so we are free to take them as we want. For both tetrahedra, we fix three normals imposing them to be opposite to the normals of the face shared with the first three tetrahedra; the last one is then fixed by the closure constraint. By construction, the directions of the last two normals obtained from the closure constraints are automatically normal, but one of the two will not necessarily be a unit vector. Imposing unit norm is the constraint reducing the 6 shape variables to the 5 degrees of freedom of a vector geometry. The constraint so obtained is highly non-linear in the areas and shapes, so that we are not even able to write it in a closed form, and very difficult to solve in order to provide only five independent variables. 

\begin{figure}[ht]
    \centering
    \includegraphics[width=5cm]{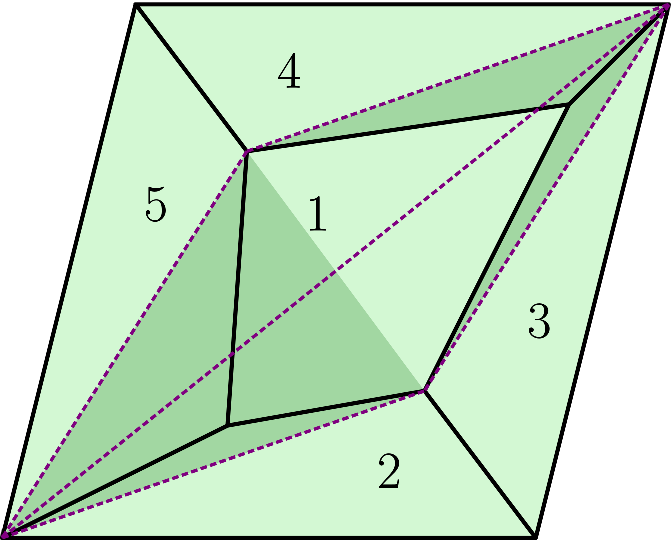}
    \caption{\label{KM1}\small{\emph{The Kapovich-Millson diagram of a vector geometry. The numbers label five parallelograms in $\R^3$, each corresponding to a tetrahedron. The internal black lines are 8 pairs $j_{ab}\vec n_{ab}=-j_{ab}\vec n_{ba}$, and the 4 external ones are opposite pairwise, and could be identified to give the whole configuration the topology of a torus. In dashed the independent diagonals that parametrise the vector geometries at fixed areas.}}}
\end{figure}

There is an alternative parametrisation that can be reduced to five independent variables, but which is not gauge-invariant. Choosing the twisted-spike gauge in which the normals are all pairwise antiparallel, we can express a vector geometry in terms of $10$ areas, $4$ three dimensional dihedral angles (1 shape parameter per tetrahedron) and $1$ extra angle between faces of different tetrahedra. To prove this statement is useful to study the problem in the Kapovich-Millson (KM) space  \cite{Kapovich}. A tetrahedron is represented in this space as a $4$-sided polygon in $\R^3$ with edge vectors $\vec v_{ab}:=j_{ab}\vec n_{ab}$.
For the twisted spike we can embed all $5$ tetrahedra in the same KM space. The orientation conditions read as an identification of sides (with pairwise opposite orientation), and can be represented as in the example of Fig.~\ref{KM1}. We can then find KM variables for the vector geometries, for instance by reconstructing all normals in terms of the five diagonals in dashed in the Figure.
To that end, we first notice that the external parallelogram $(\vec{n}_{24}$, $\vec{n}_{42}$, $\vec{n}_{35}$, $\vec{n}_{53})$ is fixed (up to a global rotation) by the two spins $(j_{24},j_{35})$ and the (squared) length of its diagonal
\begin{subequations}\label{cinese}
\be\label{uno}
\left(j_{53}\vec{n}_{53} +j_{42}\vec{n}_{42}\right)\cdot\left(j_{53}\vec{n}_{53}+ j_{42}\vec{n}_{42}\right) = j_{53}^2 + j_{42}^2 + 2  j_{53} j_{42} \cos \varphi_{53,42}.
\ee
Next, we look at the six edges spanned by this diagonal, the diagonals of the polygons $4$ and $5$, and the three vectors $\vec{v}_{53}$, $\vec{v}_{42}$ and $\vec{v}_{45}=-\vec{v}_{54}$. This six edges define a tetrahedron, hence from $j_{35}$, $j_{24}$, $j_{45}$, \Ref{uno} and the two (squared) length of the diagonals 
\be
\left(j_{54}\vec{n}_{54}+ j_{53}\vec{n}_{53}\right)\cdot\left(j_{54}\vec{n}_{54}+ j_{53}\vec{n}_{53}\right) = j_{54}^2 + j_{53}^2 + 2  j_{54} j_{53} \cos \varphi_{34,5},
\ee
\be
\left(j_{42}\vec{n}_{42}+ j_{45}\vec{n}_{45}\right)\cdot\left(j_{42}\vec{n}_{42}+ j_{45}\vec{n}_{45}\right) = j_{42}^2 + j_{45}^2 + 2  j_{42} j_{45} \cos \varphi_{25,4},
\ee
we can reconstruct the normals $\vec{n}_{53}$, $\vec{n}_{42}$ and $\vec{n}_{45}$. 
With the same logic we can determine $\vec{n}_{23}=-\vec{n}_{32}$ from the two (squared) length of the diagonals of the polygons 2 and 3, 
\be
\left(j_{32}\vec{n}_{32}+ j_{35}\vec{n}_{35}\right)\cdot\left(j_{32}\vec{n}_{32}+ j_{35}\vec{n}_{35}\right) = j_{32}^2 + j_{35}^2 + 2  j_{32} j_{35} \cos \varphi_{25,3},
\ee
\be
\left(j_{24}\vec{n}_{24} +j_{23}\vec{n}_{23}\right)\cdot\left(j_{24}\vec{n}_{24}+ j_{23}\vec{n}_{23}\right) = j_{24}^2 + j_{23}^2 + 2  j_{24} j_{23} \cos \varphi_{34,2}.
\ee
\end{subequations}

Then, the diagonal of the polygon $1$ is fixed by 3d closure as the norm of $j_{54}\vec{n}_{54}+j_{53}\vec{n}_{53}+j_{24}\vec{n}_{24}+j_{23}\vec{n}_{23}$.
Finally also by closure we determine $\vec{n}_{25}$ and $\vec{n}_{34}$. 

In the end, we have reconstructed pairwise opposite normals from      the ten areas $j_{ab}$ and five angles $\vphi$'s used in \Ref{cinese}. Out of these, 4 are dihedral angles of tetrahedra, hence invariant under local rotations. The initial one in \Ref{uno} however is an angle between normals of different tetrahedra, $\cos\varphi_{53,42}:=\vec{n}_{53}\cdot\vec{n}_{42}$, and it is not gauge invariant.

\section{Higher valence and polytopes}

The relation of the coherent $15j$ symbol to a flat 4-simplex and its Regge action is a beautiful generalisation of the Ponzano-Regge result for the $6j$ symbol, and it is natural to ask the question of whether a similar relation exists for more complicated SU(2) graph invariants, or higher-valence vertices in the language of spin foams. 
The amplitude \Ref{Av} generalises to an arbitrary graph is a straightforward manner; and so does the twisted geometry interpretation of the boundary data as a 3d collection of flat polyhedra, adjacent by faces as specified by the graph. 
The question is whether these polyhedra are geometrically glued together at the critical points, and whether the gluing defines a 4d object. As we will see, the case of the 4-simplex graph is special in two ways: first, the integrand has two distinct saddles only for shape-matched data, defining a 3d Regge geometry; second, that 3d Regge geometry admits a unique flat embedding. The first result means that the amplitude oscillates only for metric data; the second that the metric data can be given a 4d interpretation. Both properties are lost in generalisations to more complicated graphs: the twin saddle points do not suffice to identify a shape-matched 3d geometry, and even those that do, can not in general be associated with the boundary of a flat 4d object with areas given by the spins, hence a further restriction would be necessary to link the asymptotic oscillations to a Regge action. In this Section we answer these questions, and characterise what geometric objects admit twin saddles, and what special subclasses do identify a flat polytope. Only for the latter the asymptotic oscillates with the cosine of the polytope Regge action.

\subsection{Asymptotics of generalised graph invariants}
The formula for the coherent 4-simplex amplitude immediately generalises to an arbitrary graph $\G$ with $N$ nodes and $L$ links,
\be\label{AG}
A_\G(j_{ab},\vec n_{ab})= 
\int \prod_{a=1}^{N} dg_{a} \prod_{(ab)}\bra{-\vec{n}_{ab}}g_{a}^{-1}g_{b} \ket{\vec {n}_{ba}}^{2j_{ab}},
\ee
where $a,b$ label nodes as before. Again we choose the orientation of each link so that $a<b$, for bookkeeping simplicity. The boundary data are then a collection of areas $j_{ab}$ and pairs of normals $(\vec n_{ab}, \vec n_{ba})$ associated to each oriented link $(ab)$. 
For ease of language, we will refer to first-neighbour nodes also as `linked' nodes, meaning they are connected by a single link.
Expanding this formula in an orthonormal basis of graph invariants one obtains a linear combination of $nj$ symbols associated to the graph, weighted by coherent intertwiners describing each a polyhedron with the number of faces determined by the valence of the node.
The formula is valid for any graph; however the existence of two distinct saddle points will put strong restrictions on them, and we will mostly discuss the case of graphs dual to tessellations of a 3-sphere.

Extending the previous saddle point approximation is straightforward: the only new feature that appears is that not all pairs of nodes are first neighbours. The critical points for which both real and imaginary parts of the gradient of the action vanish are identified as before by vector geometries,
\be\label{vecgeogen}
\sum_{b\neq a}j_{ab}\vec{n}_{ab} = 0, \qquad R_b \vec{n}_{ba}=-R_a \vec{n}_{ab},
\ee
with the closure holding for each node and the second condition for all pairs of first-neighbour nodes $ab$. These data span $3(L-N)$ dimensions, and their characterisation can be derived generalising the procedure described in the previous Section.
To solve the critical point equations we follow our earlier derivation: we look first at pairwise anti-parallel boundary normals $\vec{n}_{ba}=- \vec{n}_{ab}$, and later use gauge invariance at the nodes to extend the results to an arbitrary configuration \Ref{vecgeogen}. 

For pairwise anti-parallel normals there exist trivially a critical point at the identity, $R_a=\Id \ \forall a$. This saddle point exists for any graph, and leads to a power-law asymptotics. The existence of a second saddle point will put more severe  restrictions on the graph. Graphs that are dual to the boundary of a polytope will turn out to always admit data leading to a second saddle point, and we will restrict our attention to them.
To look for a second saddle point, we begin as before by using the gauge freedom to fix $g_1=\Id $, and solving the equations \Ref{rotCritPoint} for all nodes $a$ and $b$ first-neighbours to 1. This 
determines the group elements
\be\label{genspher}
R_a := e^{2i\th_{1a} \vec n_{a1}\cdot \vec J},\qquad 
\cos\th_{1a} = \f{\cos\vphi^{(b)}_{1a} + \cos\vphi^{(a)}_{1b} \cos\vphi^{(1)}_{ab}}{\sin\vphi^{(a)}_{1b} \sin\vphi^{(1)}_{ab}},
\ee
now valid for all nodes first-neighbours to 1.
The spherical cosine law for $\th_{1a}$ so derived has to hold for every node $b$ linked to both 1 and $a$, else there is no solution. The geometric meaning of these generalised edge-independence conditions and of the data satisfying them will be discussed in details below, for the time being we just assume that non-trivial solutions exist. We remark that we can also determine via \Ref{sphericalcosinelaw} the angles $\th_{ab}$ for $a,b$ first-neighbours to 1, and that none of the $\th_{ab}$ angles derived so far can be zero, or else they are all zero.

Next, we pick one of the nodes not linked to 1 whose group element we still have to determine, denote it $c$, and look at its critical point equations with all $a$'s linked to 1:\footnote{We assume that there are at least 2 of them. This is the case for  graphs dual to the boundary of a polytope. If it is not true, say for instance a graph that can be reduced to two disconnected graphs by cutting a single link, then interesting degeneracies can appear in the saddle point analysis, leading to slower power law decays like for the 3d degenerate configurations, and like those, with less interesting geometric interpretation.}
\be
R_c^{-1}R_a \vec n_{ca} = -\vec n_{ac} =\vec n_{ca} .
\ee
To solve these equations we iterate the above procedure. We choose a node $x$ linked to both $c$ and 1 (let's say with $x>c$ to fix the orientation of the rotation $R_c$, if $x<c$ one should replace $R_c$ with $R^{-1}_c$ in the formulas below), and reparametrise $\tl R_c = R_x^{-1}R_c$. The critical point equations splits in two sets similar to \Ref{rotCritPoint}, that is
\be
\tilde R_c^{-1} \vec n_{c x} = \vec n_{cx}, \qquad \tilde R_c^{-1} R^{-1}_x R_a  \vec n_{c a} = \vec n_{ca}. 
\ee
From the equations on the left we immediately find the axis of rotation, $\tl R_c= e^{2i\th_{x c} \vec n_{c x}\cdot \vec J}$, with an angle $2\th_{xc}$ to determine using the equations on the right. The rotation $R^{-1}_x R_a=e^{2i\th_{x a} \vec n_{xa}\cdot \vec J}$ is already determined, and projecting as before we obtain, for $\th_{xa}\neq 0$,
\be\label{spher}
\cos\th_{x c} = \f{\cos\vphi^{(b)}_{xc} + \cos\vphi^{(b)}_{xc} \cos\vphi^{(x)}_{cb}}{\sin\vphi^{(c)}_{xb} \sin\vphi^{(x)}_{cb}}.
\ee
In other words, we have extended the result \Ref{genspher} to the new base point $x$. With the composition of rotations we can at this point identify also the rotation angle and direction of $R_c$, as well as all angles $\th_{ac}$ associated to the rotations $R_c^{-1}R_a$. 

The procedure iterates in the same way to the remaining nodes. In the end, we determine all the rotations at the non-trivial critical point in terms of normals. In the process we also established that all pairs of rotations on linked nodes have the form
\be
R_a^{-1} R_b = e^{2i\th_{ab}\vec n_{ba}\cdot \vec J},
\ee
with $\th_{ab}$ given by \Ref{spher} -- with $a$ replacing $x$. These require generalised
edge-independence conditions for all angles $\th_{ab}$, namely
\be\label{gen-edge-ind}
{\cal C}_{ab,cd} = \f{\cos\vphi^{(c)}_{ab} + \cos\vphi^{(b)}_{ac} \cos\vphi^{(a)}_{bc}}{\sin\vphi^{(b)}_{ac} \sin\vphi^{(a)}_{bc}}
-\f{\cos\vphi^{(d)}_{ab} + \cos\vphi^{(b)}_{ad} \cos\vphi^{(a)}_{bd}}{\sin\vphi^{(b)}_{ad} \sin\vphi^{(a)}_{bd}}=0,
\ee
for all $a$ and $b$ first-neighbours, $c$ first-neighbours to $a$ and $d$ to $b$. Also, we had to exclude configurations with any $\th_{ab}=0$, as this would immediately collapse the system to the single trivial solution.\footnote{The reader may wonder about polytopes with parallel 4d normals between two or more polyhedra. Let us distinguish two cases. If the parallel normals concern two non-adjacent polyhedra, their vanishing relative angle never appears in the formulas above, which only include angles between first neighbours. This case can thus be described by the boundary data and has two distinct critical points. If on the other hand the vanishing dihedral angle occurs between adjacent polyhedra, this configuration will effectively lack a second critical point. Notice however that such a polytope is exactly equivalent to a polytope whose boundary has one less polyhedron and one less face. Its geometry will then appear with two distinct critical points for the reduced graph with one less node.}
Hence, a second saddle point exists only for configurations with all $\th_{ab}\neq 0$ among linked nodes and satisfying the edge-independence conditions. We will prove in the next Section that this data set is not empty, and that the effect of \Ref{gen-edge-ind} is to impose the matching of the 2d internal angles of the faces of a twisted geometry.
Finally, notice also that we assumed 3d non-degeneracy as for the 4-simplex, meaning the closure conditions are satisfied with normals non coplanar. 
This is needed for the spherical cosine laws and the polyhedral interpretation of the normals.

To compute the leading order asymptotics of these configurations, we straightforwardly generalise the expansion of the action \Ref{Sexp} and the Gaussian approximation. This yields:
\begin{itemize}
\item Vector geometries \Ref{vecgeo}, not satisfying the angle-matching conditions: a single saddle point at the identity, and 
\begin{align}
A^{\rm LO}_\G(j_{ab},\vec n_{ab})=\left( \frac{2 \pi}{ \lambda} \right)^{\f32(N-1)} \frac{2^{N-1}}{(4 \pi)^{2(N-1)}}  \frac{1}{\sqrt{\det - H^{(0)}}}.
\end{align}
\item Angle-matched vector geometries: two distinct saddle points, and
\begin{align}\label{ALOgen}
A^{\rm LO}_\G(j_{ab},\vec n_{ab})&=\left( \frac{2 \pi}{ \lambda} \right)^{\f32(N-1)} \frac{2^{N-1}}{(4 \pi)^{2(N-1)}}  
 \left( \frac{1}{\sqrt{\det - H^{(0)}}} + \frac{e^{i 2 \lambda S_R }}{\sqrt{\det -H^{(\th)}}}\right),
 \end{align}
where
\be\label{SG}
S_{\G} = \sum_{ab} j_{ab} \th_{ab}(\varphi)
\ee
is an action defined with closure and angle-matching satisfied.
\end{itemize}
The Hessian has the same block structure \Ref{Hessian}, and we do not have a closed expression for its determinant. We did not perform numerical checks, but it is reasonable to expect that the reality condition \Ref{daje} is still satisfied, in which case \Ref{ALOgen} can also be put in cosine form as \Ref{LO2}.

As in the 4-simplex case, the results extend up to a global phase to arbitrary boundary data satisfying \Ref{vecgeogen}, and to the holomorphic definition of the vertex amplitude like \Ref{Aholo}. For more generic boundary data that are not vector geometries, there are no critical points and the amplitude is thus exponentially suppressed.

\subsection{Geometric interpretation of the twin saddles: conformal twisted geometries}\label{SecCTG}

As we have seen, generalising the asymptotic expansion of the 4-simplex amplitude to an arbitrary graph is rather straightforward. What requires more work is the geometric interpretation of the results. In particular, it is not true any longer that the complete set of data admitting two distinct saddle points describes Regge geometries, nor that the asymptotic action \Ref{SG} is a Regge action: it may have the right-looking form, but it does not have the right functional dependence to be a Regge action. In this Section, we discuss the geometric meaning of the data admitting two saddle points. In the next Section, of the asymptotic action.

In the following, we will restrict attention to graphs dual to the boundary of a polytope,
\begin{figure}[ht]
\begin{center} \includegraphics[width=4cm]{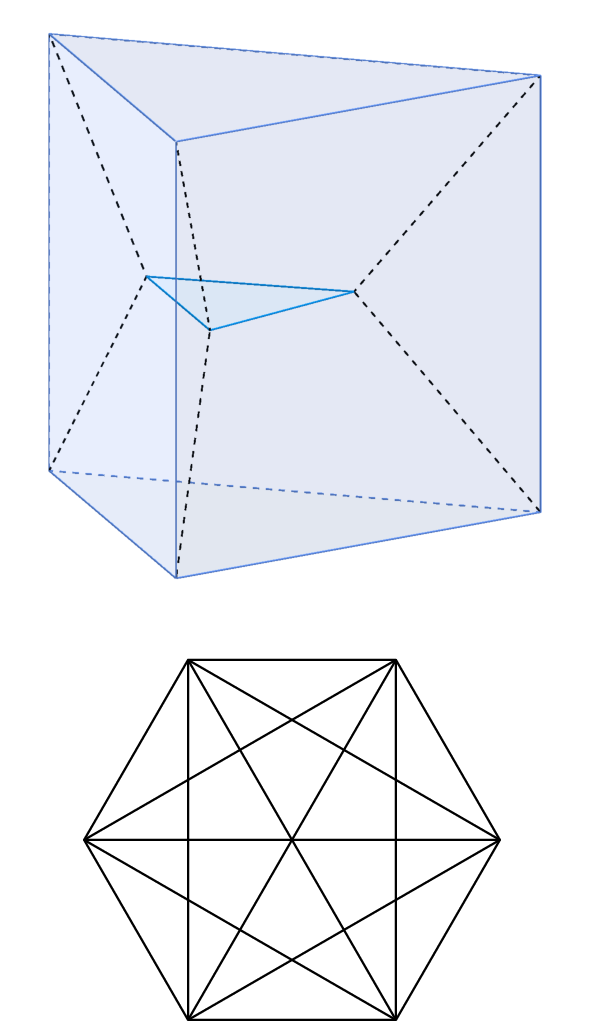} \hspace{2cm} \includegraphics[width=4cm]{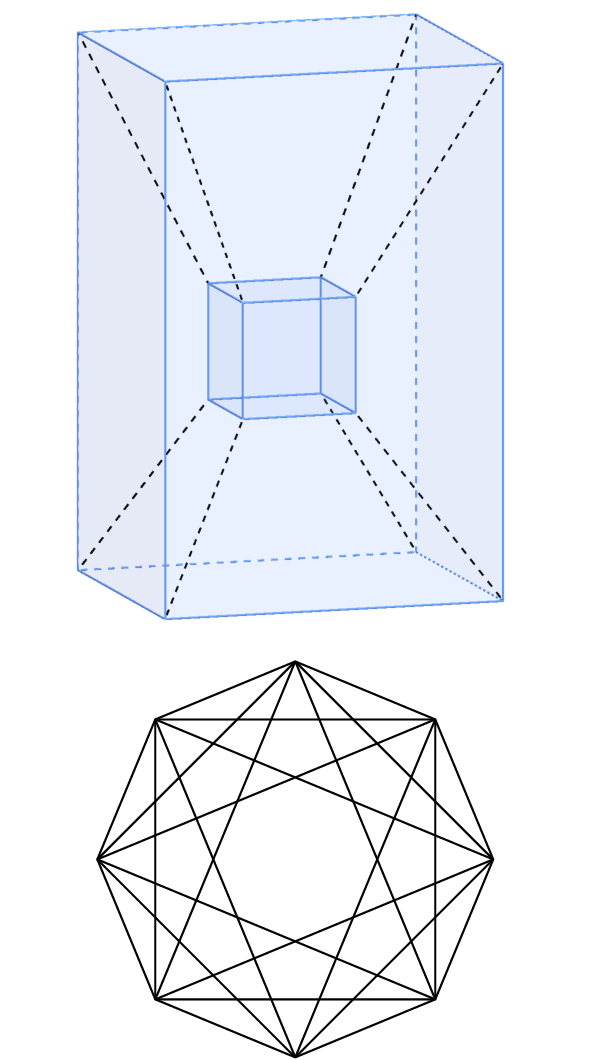} \end{center}
\caption{\label{FigSchlegels} {\small{\emph{The Schlegel diagrams of two simple polytopes and their associated boundary graphs, defined as the 1-skeleton of the dual to the boundary of the polytope. The boundary graph loses information about the polytope, for instance it is not possible to tell which of the 3-cycles of the graph identify an edge of the polytope, and which a virtual edge given by the intersection of three polyhedra's hyperplanes outside the boundary of the polytope.} Left panel: \emph{A 3-3 duo-prism, with $(V,E,F,C)=(9,18,15,6)$. The boundary is still a complete graph as in the 4-simplex, but not all 3-cycles correspond to edges. All 3-cycles identify nonetheless a generalised edge, and the spherical cosine law for the dihedral angles of the polytope correctly applies. By Minkowski theorem, an arbitrary flat convex 3-3 duo-prism has 14 degrees of freedom, which means that there exist one non-trivial flat embedding condition for its boundary 3d tessellation described by 15 edge lengths.} Right panel: \emph{A hypercuboid, with $(V,E,F,C)=(16, 32, 24, 8)$. The boundary graph is not a complete graph. In the saddle point analysis, we have spherical cosine laws only for the 3-cycles, i.e. those triples of nodes which are mutually first neighbours. By Minkowski theorem, an arbitrary convex hypercuboid has 22 degrees of freedom, thus we expect 10 non-trivial flat-embedding conditions.
 }  }} }
\end{figure}
namely dual to a tessellation $\D_3$ of the 3-sphere. See Fig.~\ref{FigSchlegels} for two examples. 
The graph alone carries of course much less information than the tessellation:
the nodes $N$ and links $L$ determine the number $C=N$ of 3d cells of the tessellation, their connectivity, and the number $F=L$ of faces; we do not know the valence of each face $f$ (namely the number of edges in its boundary, call it val$_f$), nor the
number of edges ($E$) and vertices ($V$). We can try to reconstruct this missing information from the boundary data  $(j_{ab},\vec n_{ab})$ of the coherent amplitude: the Minkowski theorem guarantees that areas and (non-coplanar) normals satisfying closure will determine locally the geometry of a flat polyhedron, and thus its adjacency matrix \cite{IoPoly}. The difficulty with this procedure is that the reconstructed twisted geometry will in general have faces of mismatched shape as well as valence: a face reconstructed as a triangle from the data on one node may be a pentagon (of the same area) using the data of the adjacent node. The mismatch prevents in general the identification of a tessellation, since we still lack an identification of the edges. This is also the case if we restrict to vector geometries \Ref{vecgeogen} with a single saddle point: generalising the procedure of Section~\ref{SecVecGeo}, one can easily convince oneself 
 that the restrictions imposed by the orientation equations are too mild, in particular mismatch in the valence of the shared faces is still allowed.
As for the dimension of the space of vector geometries, the condition of pairwise antiparallel normals means that we have a single independent normal per link; including the spins and subtracting one closure condition per node, this gives $3(L-N)$.

On the other hand, we will now prove that the configurations with two distinct critical points do allow a complete identification of the tessellation, thanks to the generalised edge-independence conditions \Ref{gen-edge-ind}: these imply the matching of the 2d dihedral angles, thus the matching of the valence of the shared faces; which in turns allows us to combinatorially identify the edges of the tessellation. 
To that end, let us start from the meaning of \Ref{spher}: this spherical cosine law gives the relation between 4d and 3d dihedral angles as defined by intersections of planes and hyperplanes. 
The difference with the 4-simplex case is that the 1d intersection of the three hyperplanes to which the polyhedra $x$, $b$ and $c$ belong needs not be a boundary of the polyhedra, but this does not affect the validity of the spherical cosine law. 
On the other hand, the solution exists only if the data satisfy the generalised edge-independence conditions \Ref{gen-edge-ind}, where we extend the use of edge to mean the triple intersection of hyperplanes, not necessarily belonging to the boundary of the polyhedron.\footnote{To help the reader pictorially with a specific example, consider a flat polytope like in Fig.~\ref{FigSchlegels}. We can compute its 4d dihedral angles using \Ref{genspher} for \emph{any} triple of pairwise-adjacent polyhedra, obtaining the same result, even if in some cases this will mean taking a triple of polyhedra that do not share an edge. E.g. in the 3-3 duo-prism computing the dihedral angle between the top and bottom prisms respect to the outside prism. \label{checaldo}}

Next, we want to show that \Ref{gen-edge-ind} imply the matching of 2d dihedral angles. This can be easily done in two steps. First, the orientation equations imply that  
the 4d dihedral angles, computed with the edge-dependent spherical cosine laws, coincide with the twist angles \Ref{defxi}:
\be
\cos\xi_{ab}^{(c)} =\frac{\left(\vec{n}_{ab}\times\vec{n}_{ac}\right)\cdot\left(\vec{n}_{ba}\times\vec{n}_{bc}\right)}{\left\Vert \vec{n}_{ab}\times\vec{n}_{ac}\right\Vert \left\Vert \vec{n}_{ba}\times\vec{n}_{bc}\right\Vert } \equiv \cos\th_{ab}^{(c)}.
\ee
Hence, \Ref{gen-edge-ind} are equivalent to the matching of twist angles
\be
\cos\xi_{ab}^{(c)} =\cos\xi_{ab}^{(d)} 
\ee
for triples of mutually first-neighbour nodes. 

\begin{figure}[ht]
\begin{center} \includegraphics[width=4.5cm]{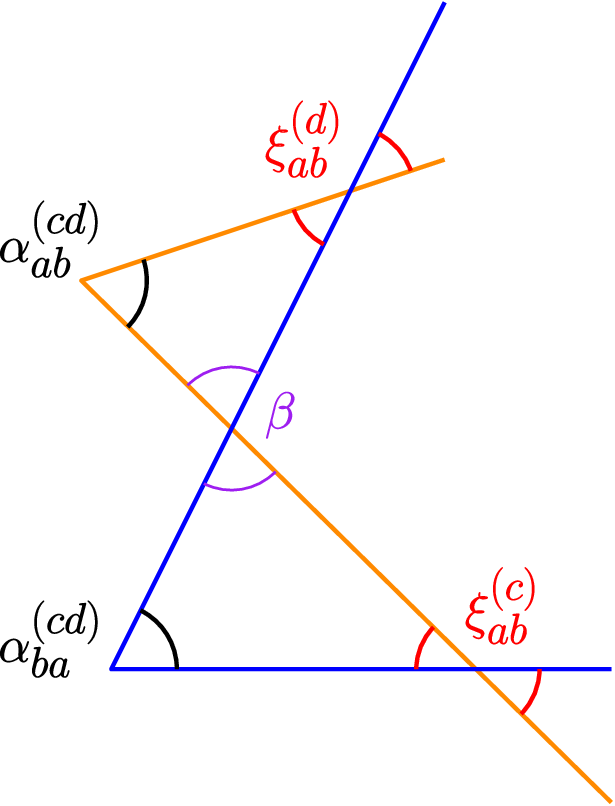}  \end{center}
\caption{\label{FigMatching} {\small{\emph{Proof that matching of twist angles implies matching of 2d dihedral angles. Here the blue and yellow lines are the edges of two adjacent polygonal faces with the same valence. The $\xi$'s are twist angles and the $\a$'s the 2d dihedral angles. More precisely, $\xi_{ab}^{(c)}$ is the twist angle between the vectors built from the data of the polyhedron $a$ and the data of the polyhedron $b$ and associated to the same edge $abc$; $\a_{ab}^{(cd)}$ and $\a_{ba}^{(cd)}$ are the 2d dihedral angle between the edges $abc$ and $abd$ computed respectively from the data of $a$ and from the data of $b$. The intersection of the faces determine two triangles, which by construction have one angle identical to one another, denoted $\b$. Then, matching of the $\xi$'s implies matching of the $\a$'s by similitude of the triangles. Notice that the edges need not be adjacent in the face, i.e. the 2d dihedral angle may not belong to the face: matching the proper internal 2d angles implies also matching the `external', or improper ones, as in higher dimensions (see footnote \ref{checaldo}).}  }} }
\end{figure}

Second, consider two adjacent polygonal faces, and pick a vertex shared by two edges, either authentic or virtual. If one of the edges is aligned between one face and the other, its twist angle is zero; it is then obvious that vanishing of the second twist angle implies matching of the 2d dihedral angles. In general, both edges may be misaligned. Then one can look at the triangles defined by the self-intersection of the faces, and easily derive that
\be
\alpha_{ab}^{(cd)} - \alpha_{ba}^{(cd)} = \xi_{ab}^{(c)} - \xi_{ab}^{(d)}.
\ee
This is a simple exercise in similarity of triangles that can be best explained with the help of Fig.~\ref{FigMatching}.
In this way one matches all the 2d dihedral angles defines by edge vectors, whether they belong or not to the polygonal face. It should be clear that the complete angle matching (of both proper and improper 2d angles) can only be only satisfied if the two polygonal faces have the same valence, as it follows trivially from the fact that the sum of the 2d angles must equal $({\rm val}_f- 2)\pi$.

\begin{figure}[ht]
\begin{center} \includegraphics[width=4cm]{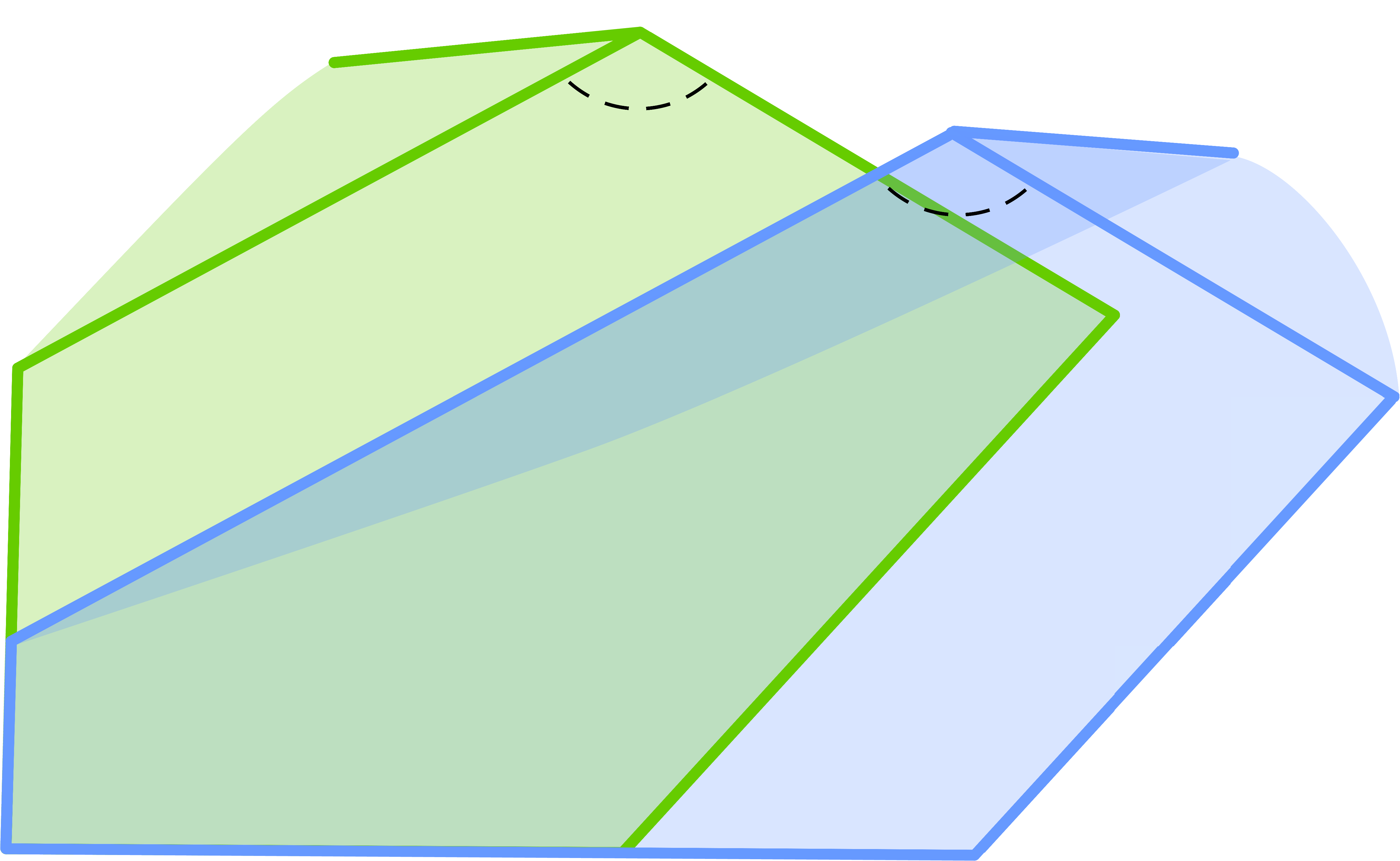}  \end{center}
\caption{\label{FigPenta} {\small{\emph{Example of faces with matched areas and angles, differing by the diagonals. The freedom of such conformal or similitude transformations is val$_f-4$. The picture also shows the shadows of the two polyhedra sharing the face, and an example of 2d angle and the 3 edges whose 3d dihedral angles are used to determine it in each polyhedron. Angle-matched configurations thus constrain the 3d dihedral angles.}  }} }
\end{figure}

This proves that the second saddle point can only exist for data that identify polyhedra adjacent by faces whose areas and angles match: an angle-matched twisted geometry. This means that the edges can be combinatorially identified. On the other hand, their lengths is not uniquely assigned: each polyhedron determines a priori a different edge length.
This is the main difference with the 4-simplex: in that case, the angle-matchings immediately imply unique edge-lengths, hence shape-matchings and the existence of a 3d Regge geometry. In the general case, adjacent polygonal faces with the 2d angles matching can still carry a conformal mismatch, whereby the area and 2d angles match, but not the diagonals, hence not the edge lengths. See Fig.~\ref{FigPenta} for an example.

The existence of saddle points for geometries with conformal mismatches at the faces was first observed in \cite{BahrSteinhaus15}, where the authors considered a hypercube graph and reduced 3d data corresponding to regular parallelepipeds (6 free spins and no free angles out of the possible 72 boundary variables). 
Our results confirm and generalise these findings. 
More recently in \cite{Bahr:2017eyi} the same authors considered truncated pyramids (3 free spins, no free angles out of 72). It was there observed that 
that the 2 saddle points in this setup collapse to one or zero depending on whether the reconstructed 4d dihedral angle from the 3d data vanishes or becomes imaginary, consistently with our analysis above.

It is not straightforward to compute the dimensionality of the space of conformal, angle-matched twisted geometries: neither the variables (angles) nor the constraints (angle-matchings) are linearly independent, which makes an explicit counting difficult. We leave this question to future research.
On the other hand, it is easy to count the dimensionality of the subset that describes 3d Regge geometries.   
For shape-matched configurations, the boundary data describe a unique cellular decomposition $\D_3$ of the 3-sphere: its intrinsic geometry is described by $E$ data corresponding to the edge lengths. 
The number of edges $E$ can then be counted as follows. In the most general case, each vertex of the tessellation will be 4-valent. Higher valence vertices require special alignments of normals and are thus measure-zero, or subdominant in the classification of \cite{IoPoly}. For a closed graph with 4-valent vertices we have $2V=E$. Then, from Euler's formula $-C+F-E+V=0$ we get $E=2(F-C)=2(L-N)$. This is the largest dimension 3d Regge data can span, and subdominant classes can always be derived when one or more edge lengths vanish.

The next question is what 4d geometry can be defined with the 3d Regge data. Since by construction our normals are outgoing and our dihedral angles always between $0$ and $\pi$, it is natural to look for a 4d convex embedding. However,  
it cannot be flat in general: a convex flat 4d polytope has $4C-10$ degrees of freedom, which is less than $E$ for a dominant class. There are thus more degrees of freedom in the 3d Regge data than in a flat 4d polytope with the same number of cells. 
The only exception is the case of the 4-simplex, where the number of degrees of freedom match, and the 3d Regge data identify a unique flat 4d object. 
This is not surprising: even in the smooth case, a 3d hypersurface cannot always be flatly embedded in Euclidean space, and the same is true for a Regge triangulation. 
Combinatorially, we count $E-(4C-10)$ flat-embedding conditions. For configurations that do not satisfy them, the data is not the boundary of flat polytope but rather of a curved 4d tessellation, made necessarily of more than one 4d building block. 

A way to count the geometric degrees of freedom of a flat, convex polytope is to use the Minkowski theorem: a collection of $n$ vectors in $\R^k$ satisfying the closure condition determines a unique flat and convex $k$-polytope up to isometries. Each vectors is normal to a $k-1$ facet of the polytope, and its norm gives the $k-1$ volume of the facet, and the degrees of freedom are $(n-1)k-k(k-1)/2$.
For $C$ vectors $V^I_a = V_a N^I_a$ in 4 dimensions ($a=1,\ldots C$, $I=1,\dots 4$), this gives as anticipated above $4C-10$: the 3d volumes of the polyhedral facets, and $3C-10$ independent dihedral angles $\th_{ab}$, defined as the angles between the outgoing 4-normals, $N^I_a N^J_b \d_{IJ} = \cos\th_{ab}$.

The theorem gives us also a criterion for the identification of flatly embeddable 3d Regge data.
To that end, notice that as soon as the angle-matching conditions \Ref{genspher} are satisfied, we have a unique notion of 4d dihedral angles $\th_{ab}(\vec n_{ab})$, hence of 4d normals $N_a^I(\vec n_{ab})$, up to a global SO(4) rotation. Furthermore, since closure is locally satisfied at the nodes, we know entirely the geometry of each polyhedron, including the volumes $V_a=V_a(j_{ab},\vec n_{ab})$. At this point we can forget about the local geometry of the polyhedra and their shape mismatches, and just use the 4d Minkowski theorem: the data $(V_a, N_a)$ 
can be used to define a unique convex polytope in $\R^4$, if we select those satisfying the 4d closure condition $\sum_a V_a N_a = 0$.
Of course, the boundary geometry of this auxiliary polytope has nothing to do with the 3d geometry of the individual polyhedra, apart from the volumes that coincide by construction. In particular, the areas $A_{ab}(V_a,N_a)$ differ from the spins $j_{ab}$. It should also be clear that infinitely many angle-matched data reconstruct the same auxiliary polytope. Then, a criterion for flat embedding is to demand that the 3d Regge geometry and the auxiliary flat polytope match. That means imposing the matching of the areas, but also the 4d closure condition, which is not a priori satisfied by the 3d Regge data. Boundary data $(j_{ab},\vec n_{ab})$ satisfying closure, shape-matching conditions and furthermore
\be\label{4drec}
A_{ab}\big(V_a(j_{ab},\vec n_{ab})\big) \equiv j_{ab}, \qquad \sum_a V_a(j_{ab},\vec n_{ab}) N_a(j_{ab},\vec n_{ab}) = 0,
\ee
describe the boundary of a flat convex polytope, and can thus always be flatly embedded. The criterion \Ref{4drec} is a non-linear system of equations that we do not develop explicitly here. For the interested reader, we point out that it can be obtained from a 4d version of the adapted Lasserre algorithm developed in \cite{IoPoly}.
Accordingly, we refer to 3d Regge data satisfying these further conditions as \emph{polytope data}, a set of $4C-10=4N-10$ dimensions.

Summarising, the boundary data admitting two distinct critical points describe in general a collection of polyhedra with areas and angles of adjacent faces matching, but not complete shape matchings. 3d Regge and polytope data are special subsets. Hence, the cosine oscillatory asymptotic behaviour \Ref{ALOgen} is more general than for 3d Regge geometries, and its frequency \Ref{SG} is in general not a Regge action, but a more general action whose independent variables describe the angle-matched conformal twisted geometries. We will discuss in the next Section some properties of this generalised Regge action.

\subsection{The Regge action for polytopes}\label{SecReggePoly}

Having classified the various geometries described by the boundary data and relevant subsets, let us come back to the oscillatory behaviour of the amplitudes. The case we are interested in is when two distinct critical points exist, for which we found a relative phase proportional to the action \Ref{SG}, namely an area-angle action with closure and 2d angle-matching constraints:
\be\label{SG1}
S_\G[j_{ab}, \vphi^{(a)}_{bc},\l_{a,b} ,\m_{ab,cd}] = \sum j_{ab} \th_{ab}(\vphi) + \sum \l_{a,b} C_{a,b}(j,\vphi) + \sum \m_{ab,cd} \, {\cal C} _{ab,cd}(\vphi),
\ee
with $(\l_{a,b} ,\m_{ab,cd})$ Lagrange multipliers. See \Ref{phiclosure} and \Ref{gen-edge-ind} for the explicit forms of the constraints.

From the discussion in the previous Section, we know that the independent variables in this action span angle-matched twisted geometries, a generalisation of Regge geometries with conformal mismatch allowed. These geometries can also be mapped (infinite-to-1) to an auxiliary flat convex polytope with $\th_{ab}(\vphi)$ as dihedral angles, but whose areas are not given by $j_{ab}$. 
Hence, \Ref{SG1} is not a proper Regge action for all the geometric configurations for which it is defined, but only for a subset of them.

Although not a proper Regge action in general, \Ref{SG1} has some interesting properties thanks to the fact that the 4d dihedral angles are well defined. First, it shares the property of the Regge action of being the discretisation of the boundary's extrinsic curvature. Second, it can be consistently added up to define a bulk action for a tessellated spacetime $\D_4$, similarly to the way the 4-simplex boundary Regge action is added up to give the Regge action of a triangulation \cite{Hartle:1981cf}. Consider a combinatorial tessellation of spacetime by 4d polytopes $\s$, and endow it with areas and 3d angles satisfying closure and angle matching ($f$ labels faces and $c$ the polyhedral 3d cells). Locally the geometry of each polyhedron is well defined, hence the gluing of 4d cells is automatic. Furthermore, the deficit angles are uniquely defined,
\be\label{deficit}
\eps_f(\vphi) = 2\pi - \sum_{\s\in f} \th^\s_f(\vphi),
\ee
and thus a notion of spacetime curvature a la Regge.
Adding up the individual building blocks \Ref{SG1} we obtain an area-angle action for  $\D_4$ with closure and angle-matching,
\begin{align}
\label{SD4}
S_{\D_4}[j_f, \varphi_{ff'}^c,\l_c^f ,\m_{cc'}^{dd'}] &:= -\sum_\s S_{\G(\s)} [j, \vphi,\l,\m] \nn\\
&= \sum_{f} j_f \eps_f(\varphi) - \sum_{c, f\in c} \l_c^f C_c^f(j,\varphi) - \sum_{cc'dd'} \m_{cc'}^{dd'} C_{cc',dd'}(\varphi) -2\pi\sum_f j_f.
\end{align}
Here we have switched to a notation $j_f$ for the spins, and $\vphi_{ff'}^c$ for the 3d angles between faces $f$ and $f'$ within the polyhedron $c$, and similarly relabelled the constraints and Lagrange multipliers. The last term has no effect on the equations of motion, and is eliminated by the face weights $(-1)^{2\sum_f j_f}$ present in the BF partition function. 

Because of the conformal mismatch, the space of solutions of the equations of motion of this action is certainly larger than that of Regge calculus. It would be interesting for future work to study its geometric meaning, and whether it admits a well-defined continuum limit.
Consider then a restriction to 3d Regge data, adding by hand the remaining shape-matching constraints to \Ref{SG1}. The first term of the action now properly describes the extrinsic curvature of a 3d Regge tessellation, but not necessarily flatly embeddable.
Hence, as we vary the action restricted to 3d Regge data, we span all possible flat polytopes and also a finite amount of allowed bulk curvature. To properly describe the curvature, one should subdivide the boundary graph and the polytope into respectively tetrahedra and 4-simplices with internal faces.
One can also take this as the building block of an action for $\D_4$ like for \Ref{SD4}, and again we expect to get a theory more general than Regge calculus, with geometric meaning and continuum limit to be explored.

Finally, let us consider restricting to polytope data, adding further the flat-embedding conditions to \Ref{SG1}. Since the metric variables of a flat polytope are volumes and 4d normals, \Ref{SG1} is now equivalent to
\be\label{Rpoly}
S_\s(V_a, \th_a) := \sum_{(ab)} A_{ab}(V,\th) \th_{ab},
\ee
which is the proper Regge action one would write for a flat convex 4d polytope. Hence, for polytope data the asymptotics of the coherent amplitude \Ref{AG} oscillates with frequency given by the polytope Regge action, providing a proper extension of the result of \cite{BarrettSU2} to $nj$ symbols and polytopes.

Now one could also entertain the idea of using a polytope tessellation of 4d spacetime, and write a Regge action for it adding up the individual contributions \Ref{Rpoly}, as in the simplicial case. Again the deficit angles \Ref{deficit} are well defined and so is the Regge curvature. However, the gluing of polytopes is not automatic anymore. In fact, the geometry of each polyhedron is in general determined by the $(V_a,N_a)$ data of the whole polytope; two polytopes sharing a polyhedron will provide the same volume, but different shapes in general. To consistently add up the polytope terms to a $\D_4$ action we would need to include polyhedral shape-matching conditions. This is the 4d generalisation of the same problem of shape mismatch for twisted geometries, and it is analogue of the non-local area constraints needed for area Regge calculus.

Summarising, we can distinguish five classes of data for a coherent vertex amplitude \Ref{AG} on a general simple graph, and different associated asymptotic behaviours. In order of specialisation, these are:

\begin{itemize}
\item Twisted geometries.\footnote{With $\bar\xi_{ab}=0$ since we are working with sharp areas, see footnote \ref{footxi}. The most general twisted geometry has $6(L-N)$ dimensions.} These are data required only to satisfy closure, and span a space of dimensions $5L-6N=L+\sum_c 2({\rm val}_c-3)$, which can be parametrised by a areas $j_{ab}$ and $2({\rm val}_c-3)$ shape variables per tetrahedron with val$_c$ faces. The faces are mismatched both in shape and valence, and there is no clear notion of edges of the dual tessellation (or cellular decomposition). 
The amplitude has no saddle points and decades exponentially.

\item Vector geometries. The data satisfy also the orientation equations, namely the normals are pairwise anti-parallel up to local rotations. These data span a space of dimensions $3(L-N)$, for which the shapes and valences of faces are still mismatched. A parametrisation can be found generalising the procedure of Section \ref{SecVecGeo}. These data can be given on any graph, the corresponding amplitude has a saddle point, and decays with a power-law.

\item Angle-matching vector geometries. The data satisfy also the edge-independence conditions \Ref{gen-edge-ind}. They describe piecewise-flat geometries that allow a conformal mismatch in the polygonal faces. The amplitude has two distinct saddle points, whose interference produces oscillations whose frequency is given by the generalised Regge action \Ref{SG1}.

\item Regge data. All shapes match, and the data describe a 3d tessellation of the 3-sphere which is uniquely characterised by the edge lengths, generically $E=2(L-N)$. Two distinct saddle points. Oscillations with a Regge action giving the extrinsic curvature of a generically curved bulk.

\item Polytope data. 3d Regge data admitting a flat embedding, a set of $4N-10$ dimensions. Two distinct saddle points, oscillations with a proper Regge action \Ref{Rpoly} corresponding to the extrinsic geometry of a flat convex 4d polytope.

\end{itemize}

For completeness, we should add that we excluded from our analysis degenerate configurations with vanishing 3d volumes, for which the Hessian can be degenerate thus giving slower power-law fall offs and dominating amplitudes, but whose geometric interpretation is less interesting.

\section{Implications for spin foam models of quantum gravity}\label{SecQG}
Although our background and motivations come from quantum gravity, the results presented concern purely algebraic and geometric properties of SU(2) invariants, and are valid in any context in which they enter. In this Section we would like to discuss their specific relevance to spin foam models of quantum gravity \cite{PerezLR}, aimed at readers with a certain familiarity with this approach.

The current state of the art in spin foam quantum gravity is exemplified by the EPRL model \cite{EPRL}. A key property of this model is the emergence of the Regge action in the asymptotics of its vertex amplitude \cite{BarrettEPRasymp,BarrettLorAsymp}, since it makes a connection with general relativity in the semiclassical limit conceivable. From this point of view, it came as a bit of a surprise when \cite{BarrettSU2} showed that the Regge action appears also in the asymptotics of the $15j$ symbol, which is the vertex amplitude for a spin foam model of SU(2) topological BF theory \cite{Ooguri:1992eb}. But there is no tension here: at the level of a single 4-simplex both models are 4d flat, and whether or not one gets the Regge behaviour depends on the boundary data, which can be arbitrarily chosen in both cases.\footnote{The only difference in the Regge behaviour is that for the EPRL model the Immirzi parameter shows up in front of the action, consistently with the area spectrum of loop quantum gravity. This is an immediate consequence of the linear embedding of SU(2) irreps in SO(4) or $\SL(2,\C)$ ones imposed by the primary simplicity constraints. These constraints do not restrict in any way the 4-simplex boundary data, which can still be arbitrarily chosen.}
Crucial difference between the two models is expected to show up when 4-simplices are glued together to form a triangulated 4-manifold: BF theory is invariant under Pachner moves (once it is suitably regularized), whereas the EPRL model is not.
A key open question is whether this difference is enough to make the EPRL model capture the correct Regge dynamics on the full triangulation, something that BF theory does not do. A difficulty in this respect is associated with properly treating the variation with respect to the bulk spins, leading to a possible flatness problem and an ongoing open debate \cite{Bonzom:2009hw,HellmannFlatness,AlexandrovSimplClosure,RovelliMagic,HanZhangLor,Collet,Oliveira:2017osu}. A more detailed understanding of the dynamics on a full triangulation is needed, and this is one of our goals. The results presented here can be seen as preliminary work in that direction, useful in various ways.

First of all, the interpretation of the boundary data (twisted, vector, Regge geometries)  is exactly the same in BF theory and in the EPRL model, as is the boundary Hilbert space. More importantly, the structure of the coherent vertex amplitude \Ref{Av} is the same, the only difference being that the SU(2) matrices are replaced by SO(4) or $\SL(2,\C)$ unitary matrices in one-to-one correspondence with SU(2) matrices -- via the Immirzi-parameter-dependent $Y$-map; hence, also the action \Ref{15SU2action} for the saddle point approximation has the same structure, and we can learn a great deal about the EPRL asymptotics by simply looking at SU(2) vertex amplitudes.\footnote{For the Euclidean case the action is the sum of two actions like \Ref{15SU2action}, corresponding to the splitting $\so(4)\simeq\su(2)\oplus\su(2)$. For the Lorentzian case there is a technical difference, as the infinite dimensionality of the unitary matrices requires the intermediate use of spinors instead of 3-vectors, so some work is needed to adapt our approach. Let us also point out that Lorentzian amplitudes can be exactly written as linear combinations of SU(2) vertex amplitudes weighted by 1-dimensional boost integrals at the edges \cite{Boosting}. In this case the results here obtained are directly applicable to the Lorentzian EPRL model, but they need to be completed with the analysis of the large spin limit of the booster functions \cite{Pierre}. This approach will be discussed in more details in \cite{noiLor}.}
We then expect the results presented in Sections 2-4 to extend in a simple way to the EPRL model, as we will show and make use of in future publications.
In particular the better understanding of the precise link between the saddle point conditions and the constraints of area-angle Regge calculus, and the fact that there aren't higher order corrections to the Regge action, both relevant when studying the dynamics on a full triangulation. 

Secondly, a generalized (i.e. non-simplicial) EPRL vertex amplitude for arbitrary graphs was introduced in \cite{KKL,CarloGenSF}, and it allows the model to provide transition amplitudes for arbitrary spin networks (when restricting to simplicial ones, one redefines the Hilbert space of LQG in terms of 4-valent spin networks only). It is unknown, and important to establish, whether the generalized amplitude still shows a large spin limit compatible with a discretization of general relativity, this time on a cellular decomposition instead of a triangulation. For that, one has to generalize Barrett's analysis \cite{BarrettEPRasymp,BarrettLorAsymp} to non-simplicial vertices. Our results presented in Section 5 for the simpler case of SU(2) show a pathway for doing this. Again the structure of the vertex amplitude for non-simplicial EPRL model, which is identical to our \Ref{AG} but with SO(4) or $\SL(2,\C)$ matrices now, strongly suggests that the result summarized in the end of Section \ref{SecReggePoly} and in Table~\ref{TablePolytopes} will be the same, or very similar; and this is further supported by the consistency of our results with the observations made in \cite{BahrSteinhaus15} 
for the Euclidean EPRL-KKL model. 
Inspection of the critical point equations of the EPRL-KKL model shows that the analysis presented here can be straightforwardly applied. Additional work is required for the Lorentzian case.
If the emergence of conformal twisted geometries from generalized vertices is a generic feature of these models,
there is a clear conclusion that can be drawn from our results: either one can show that the generalized Regge action \Ref{SG1} discussed at length in the previous Section is a good discretization of general relativity, or the model as it is has a worse behaviour with non-simplicial vertices, which may be problematic for the semiclassical limit.\footnote{To be fair, the negative answer would not be unexpected: as proved already in the defining paper \cite{EPR}, the linear simplicity constraints used in the model implement all of the classical ones only thanks to the use of 4-simplices. Our discussion suggests that for non-simplicial foams one may want to consider additional constraints amenable to select the data leading to \Ref{Rpoly}. Alternative considerations to impose additional constraints in the non-simplicial EPRL model have recently appeared in \cite{Bahr:2017ajs}, with the same motivations.} 

As a final comment, we have for the sake of concreteness focused our discussion on the EPRL model. But our results have direct applications to the spin foam formalism in general, along the lines described above, as long as the boundary Hilbert space and strategy to define the vertex amplitudes are the same.

\section{Conclusions}\label{SecConcl}
In the paper we presented two different types of results. Numerical results, aimed at studying the accuracy of the saddle point approximation of \cite{BarrettSU2}; and analytic, aimed at extending it to more general SU(2) graph invariants. The numerics show that the accuracy is very high already at low spins, and give insights into the higher-order corrections: the frequency of oscillations and the global phase appear to be numerically exact to all orders. The analytic results show that the asymptotic relation between SU(2) invariants and 4d Euclidean geometric objects extends to more general graphs. There exist saddle points for special configurations of the boundary data, which have an interesting geometric interpretations. The necessary condition for having at least one saddle point is a straightforward generalization of the notion of vector geometries already introduced by Barrett, a subset of the most general twisted geometry boundary data. The subset admitting two different saddles, which carries the most interesting cosine asymptotic behaviour, is on the other hand more subtle: it selects a class of conformal twisted geometries with 2d angle matchings. For the 4-simplex case this automatically implies a 3d Regge geometry, but not so for more general graphs. The frequency of the cosine oscillations is thus given by a generalisation of the Regge action, whose structure and meaning we briefly discussed, suggesting new directions to explore. A subset of the twin saddles describes 3d polyhedral Regge geometries, which in general cannot be flatly embedded: these are boundary data for curved bulk discretisations. We provided also a criterion for the flat embedding, identifying reduced data compatible with the boundary of a 4d flat convex polytope. In this case the asymptotic action coincides with the 4d Regge action defined as the total extrinsic curvature of the polytope.

Our results apply to SU(2) graph invariants and BF theory, but are also relevant to quantum gravity models based on constrained BF theory, like the EPRL model. The numerical study of Barrett's  asymptotic formula for the Lorentzian EPRL model will appear in a companion paper \cite{noiLor}. Future related work is to investigate the implications and extensions of our analytic results to the EPRL model with both simplicial and generalised vertices, and study the problem of asymptotics on extended 4d triangulations, for which the relation to solutions of the Regge equations is subject of a crucial debate \cite{Bonzom:2009hw,HellmannFlatness,AlexandrovSimplClosure,RovelliMagic,HanZhangLor,Collet,Oliveira:2017osu}. 
It would also be interesting to explore if our techniques can be generalised to quantum deformations, see e.g. \cite{Haggard:2015yda,Dupuis:2014fya}.

\bigskip

\emph{Node added after publication:} Upon further investigations, we realized that the way the angle-matching conditions arise on a general graph is slightly more subtle than how it is described here, and requires also looking at projections involving higher cycles of the graph. A more detailed and extensive analysis appears in \cite{NoiDalFuturo}. While this paper deals with the group  $\SL(2,\C)$, the considerations on the emergence of angle-matching conditions are identical and directly applicable to the SU(2) case treated here. This more detailed analysis also exposed the possibility of more than two critical points for a particular class of graphs, as firstly pointed out in \cite{Ben}. Apart from this novelty, the main conclusions are unchanged.

\section*{Acknowledgments}
We are thankful to Marios Christodoulou and Fabio D'Ambrosio for useful exchanges on coding $nj$-symbols on Mathematica, to Rafael Sorkin for discussions on flat embeddings of polytopes, and to Robin Reuben and Hal Haggard for many discussions and comments on a draft of the paper. The work of Pietro was supported by the Templeton foundation. Marco acknowledges support from the Scuola Normale Superiore di Pisa and its Erasmus program, and Giorgio from the Erasmus program of Turin University. 

\appendix

\section{Boundary data for geometric 4-simplices: an explicit example}\label{AppA}
We give here the explicit example of the reconstruction of the twisted spike for an equilateral 4-simplex with all areas $A$.
We start from its definition as the convex hull of five vertices in $\R^4$, which we take to be
\begin{center}
\begin{tabular}{ l l  }
$v_1$: & $\sqrt{A}\left(0,3^{-3/4},3^{-1/4},2^{-1/2}3^{-3/4}\right)$\\
$v_{2}$:&$\sqrt{A}\left(0,3^{-3/4},-3^{-1/4},2^{-1/2}3^{-3/4}\right)$ \\
$v_{3}$:&$\sqrt{A}\left(0,-2\cdot3^{-3/4},0,2^{-1/2}3^{-3/4}\right)$ \\
$v_{4}$:&$\sqrt{A}\left(0,0,0,-2^{-1/2}3^{1/4}\right)$ \\ 
$v_{5}$: & $\sqrt{A}\left(-5^{1/2}2^{-1/2}3^{-1/4},0,0,0\right)$ \\ 
\end{tabular}
\end{center}
We define the 4d normals to the tetrahedra as (normalised) external products of triples of edge vectors, and then orienting them to be outgoing:
\begin{center}
\begin{tabular}{ l l l }
$\tau_{1}$&$\{v_{1},v_{2},v_{3},v_{4}\}:
 $&$ N_{1}=\left(1,0,0,0\right)$ \\
$\tau_{2}$&$\{v_{1},v_{2},v_{3},v_{5}\}:
 $&$ N_{2}=\left(\frac{1}{4},0,0,-\frac{\sqrt{15}}{4}\right)$ \\
$\tau_{3}$&$ \{v_{1},v_{2},v_{4},v_{5}\}:
 $&$ N_{3}=\left(\frac{1}{4},-\sqrt{\frac{5}{6}},0,\frac{1}{4}\sqrt{\frac{5}{3}}\right)$ \\
$\tau_{4} $ &$\{v_{1},v_{3},v_{4},v_{5}\}:
 $&$N_{4}=\left(\frac{1}{4},\frac{1}{2}\sqrt{\frac{5}{6}},-\frac{1}{2}\sqrt{\frac{5}{2}},\frac{1}{4}\sqrt{\frac{5}{3}}\right)$ \\
 $\tau_{5} $& $\{v_{2},v_{3},v_{4},v_{5}\}:
 $& $N_{5}=\left(\frac{1}{4},\frac{1}{2}\sqrt{\frac{5}{6}},\frac{1}{2}\sqrt{\frac{5}{2}},\frac{1}{4}\sqrt{\frac{5}{3}}\right)$\\ 
\end{tabular}
\end{center}
We then compute the SO(4) rotations $G_a$ mapping $N_a$ to $N_1$.
These act on the plane $\{N_1,N_a\}$ by an angle $\th_{1a}=\arccos(-1/4)$. Now the vertices lie all in 3d, and we can compute the 3d normals from the (normalised) external products of the edge vectors. The configuration defines the spike of the left panel of Fig.\ref{FigSpikes}. To get the twisted spike of the right panel, we apply a 3d rotation around each normal $\vec n_{a1}$ of an angle $\arccos(-1/4)$. The resulting configuration of 3d normals is reported in the following table.
\begin{center}
\resizebox*{1\textwidth}{!}{
\begin{tabular}{| c | c c c c c  }
\hline 
$\vec n_{ab}$ & $ {1}$  & $ {2}$  & $ {3}$  & $ {4}$  & $ {5}$ \tabularnewline
\hline 
$ {1}$  &  & $\left(0,0,1\right)$ & $\left(\frac{2\sqrt{2}}{3},0,-\frac{1}{3}\right)$ & $\left(-\frac{\sqrt{2}}{3},\sqrt{\frac{2}{3}},-\frac{1}{3}\right)$ & $\left(-\frac{\sqrt{2}}{3},-\sqrt{\frac{2}{3}},-\frac{1}{3}\right)$\tabularnewline
$ {2}$  & $-\left(0,0,1\right)$ &  & $\left(-\frac{1}{3\sqrt{2}},-\sqrt{\frac{5}{6}},\frac{1}{3}\right)$ & $\left(\frac{1-3\sqrt{5}}{6\sqrt{2}},\frac{\sqrt{3}+\sqrt{15}}{6\sqrt{2}},\frac{1}{3}\right)$ & $\left(\frac{1+3\sqrt{5}}{6\sqrt{2}},\frac{\sqrt{5}-1}{2\sqrt{6}},\frac{1}{3}\right)$\tabularnewline
$ {3}$  & $-\left(\frac{2\sqrt{2}}{3},0,-\frac{1}{3}\right)$ & $-\left(-\frac{1}{3\sqrt{2}},-\sqrt{\frac{5}{6}},\frac{1}{3}\right)$ &   & $\left(\frac{3+\sqrt{5}}{6\sqrt{2}},\frac{1-\sqrt{5}}{2\sqrt{6}},\frac{\sqrt{5}}{3}\right)$ & $\left(-\frac{\sqrt{5}-3}{6\sqrt{2}},-\frac{1+\sqrt{5}}{2\sqrt{6}},-\frac{\sqrt{5}}{3}\right)$\tabularnewline
$ {4}$  & $-\left(-\frac{\sqrt{2}}{3},\sqrt{\frac{2}{3}},-\frac{1}{3}\right)$ & $-\left(\frac{1-3\sqrt{5}}{6\sqrt{2}},\frac{\sqrt{3}+\sqrt{15}}{6\sqrt{2}},\frac{1}{3}\right)$ & $-\left(\frac{3+\sqrt{5}}{6\sqrt{2}},\frac{1-\sqrt{5}}{2\sqrt{6}},\frac{\sqrt{5}}{3}\right)$ &   & $\left(\frac{1}{3}\sqrt{\frac{5}{2}},\frac{1}{\sqrt{6}},-\frac{\sqrt{5}}{3}\right)$\tabularnewline
$ {5}$  & $-\left(-\frac{\sqrt{2}}{3},-\sqrt{\frac{2}{3}},-\frac{1}{3}\right)$ & $-\left(\frac{1+3\sqrt{5}}{6\sqrt{2}},\frac{\sqrt{5}-1}{2\sqrt{6}},\frac{1}{3}\right)$ & $-\left(-\frac{\sqrt{5}-3}{6\sqrt{2}},-\frac{1+\sqrt{5}}{2\sqrt{6}},-\frac{\sqrt{5}}{3}\right)$ & $-\left(\frac{1}{3}\sqrt{\frac{5}{2}},\frac{1}{\sqrt{6}},-\frac{\sqrt{5}}{3}\right)$ & \tabularnewline
\end{tabular}
}
\end{center}

\providecommand{\href}[2]{#2}\begingroup\raggedright\endgroup

\end{document}